\documentclass[aps,twocolumn,superscriptaddress]{revtex4-2}

\usepackage{graphicx}
\usepackage{amssymb,amsmath}
\usepackage{braket}

\newcommand{\Bi}{$^{209}$Bi}
\newcommand{\Hethree}{$^{3}$He}
\newcommand{\Sitwoeight}{$^{28}$Si}
\newcommand{\Sitwonine}{$^{29}$Si}
\newcommand{\sqrtHz}{$\sqrt{\text{Hz}}$}

\begin{document}
\title{In-situ amplification of spin echoes within a kinetic inductance parametric amplifier}

\author{Wyatt Vine}
\thanks{These two authors contributed equally}
\affiliation{School of Electrical Engineering and Telecommunications, UNSW Sydney, Sydney, NSW 2052, Australia}
\author{Mykhailo Savytskyi}
\thanks{These two authors contributed equally}
\altaffiliation{Present address: IQM Finland Oy, Espoo 02150, Finland}
\affiliation{School of Electrical Engineering and Telecommunications, UNSW Sydney, Sydney, NSW 2052, Australia}
\author{Daniel Parker}
\affiliation{School of Electrical Engineering and Telecommunications, UNSW Sydney, Sydney, NSW 2052, Australia}
\author{James Slack-Smith}
\affiliation{School of Electrical Engineering and Telecommunications, UNSW Sydney, Sydney, NSW 2052, Australia}
\author{Thomas Schenkel}
\affiliation{Accelerator Technology and Applied Physics Division, Lawrence Berkeley National Laboratory, Berkeley, California 94720, USA}
\author{Jeffrey C. McCallum}
\affiliation{School  of  Physics,  University  of  Melbourne,  Melbourne,  Victoria  3010,  Australia}
\author{Brett C. Johnson}
\affiliation{Centre of Excellence for Quantum Computation and Communication Technology, School of Engineering, RMIT University, VIC, 3001, Australia}
\author{Andrea Morello}
\affiliation{School of Electrical Engineering and Telecommunications, UNSW Sydney, Sydney, NSW 2052, Australia}
\author{Jarryd J. Pla}
\email[]{jarryd@unsw.edu.au}
\affiliation{School of Electrical Engineering and Telecommunications, UNSW Sydney, Sydney, NSW 2052, Australia}

\date{\today}

\begin{abstract}
The use of superconducting micro-resonators in combination with quantum-limited Josephson parametric amplifiers has in recent years lead to more than four orders of magnitude improvement in the sensitivity of pulsed Electron Spin Resonance (ESR) measurements. So far, the microwave resonators and amplifiers have been designed as separate components, largely due to the incompatibility of Josephson junction-based devices with even moderate magnetic fields. This has led to complex spectrometers that operate under strict environments, creating technical barriers for the widespread adoption of the technique. Here we circumvent this issue by inductively coupling an ensemble of spins directly to a weakly nonlinear microwave resonator, which is engineered from a magnetic field-resilient thin superconducting film. We perform pulsed ESR measurements with a $1$~pL effective mode volume and amplify the resulting spin signal using the same device, ultimately achieving a sensitivity of $2.8 \times 10^3$ spins in a single-shot Hahn echo measurement at a temperature of 400 mK. We demonstrate the combined functionalities at fields as large as 254~mT, highlighting the technique's potential for application under more conventional ESR operating conditions.
\end{abstract}

\maketitle

\section{Introduction}

Electron Spin Resonance (ESR) spectroscopy is a technique used throughout Physics, Biology, Chemistry and Medicine to study materials through their paramagnetic properties~\cite{Schweiger2001}. To detect ESR, conventional spectrometers use a cavity to capture the weak microwave signal that is induced by the transverse magnetization of an ensemble of spins precessing in an external magnetic field. The magnitude of this signal is determined by the number of resonant spins coupled to the cavity ($N$), the spin-cavity coupling strength ($g_0$) and the quality factor ($Q$) of the cavity. The coupling strength $g_0$ depends on the degree of confinement of the magnetic energy in the microwave mode, $g_0 \propto V_m$, where $V_m$ is the magnetic mode volume \cite{Haroche2006}. Conventional ESR spectrometers utilize three-dimensional microwave cavities where $V_m \propto \lambda^3$, with $\lambda$ the wavelength of the resonant mode. An alternative approach is to use micro-resonator circuits, where the modes are confined in quasi-one-dimensional structures, such that $V_m \ll \lambda^3$ and $g_0$ is considerably enhanced \cite{morton2018,abhyankar2020}. Constructing these mirco-resonator circuits from superconducting materials also allows them to achieve high quality factors, which further enhances the spin sensitivity \cite{morton2018,artzi2022}.
 
The superconducting circuit resonator is just one tool from the field of circuit Quantum Electrodynamics (cQED) that has recently been applied to ESR. Josephson Parametric Amplifiers (JPAs) \cite{Yamamoto2008} have also been integrated into custom ESR spectrometers and used to push detection sensitivities to the quantum limit \cite{Bienfait2016a,Eichler2017}, where the noise in the measurement of a spin ensemble is set by vacuum fluctuations of the electromagnetic field. The combination of high-$Q$ superconducting microwave resonators and JPAs has ultimately resulted in sensitivities as low as 120~spins for a single Hahn echo measured at 10~mK, corresponding to an absolute sensitivity of 12~spins/$\sqrt{\mathrm{Hz}}$ \cite{Ranjan2020}.

Another approach taken to improve ESR measurement sensitivity has been to couple the spin ensemble directly to non-linear circuits. An early example of this involved coupling an ensemble of NV-centers in diamond to a superconducting transmon qubit, achieving a sensitivity of $10^5$~spins/\sqrtHz\ \cite{Kubo2012}. This was recently improved to 20~spins/\sqrtHz\ using a flux-qubit read out via an on-chip Josephson bifurcation amplifier \cite{Budoyo2020}. However, both of these approaches operate in a mode more analogous to continuous wave ESR spectroscopy. The efforts also mirror recent directions in the superconducting qubit community to engineer high efficiency measurements by reducing or eliminating the insertion loss between the system being measured and the first cryogenic amplifier \cite{Rosenthal2021,Lecocq2021}.
 
Despite these successes, previous works have all relied on technologies utilizing Josephson junctions, which are not magnetic field-resilient. This restricts the use of on-chip detection methods to low magnetic fields. Higher magnetic fields can be applied when using JPAs off-chip, where they are contained in magnetically-shielded boxes and connected to the ESR cavity via coaxial cables and microwave circulators. The disadvantage of this approach is that the extra components introduce additional insertion loss, which inevitably attenuates the spin signals before they are amplified. Moreover, JPAs typically display gain compression for input powers $\gtrsim -110~$dBm \cite{Planat2019}, which leads to signal distortion and limits the power that can be detected in pulsed ESR experiments.

In this work we demonstrate that these limitations can be overcome by using a Kinetic Inductance Parametric Amplifier (KIPA) \cite{Parker2021} coupled directly to an ensemble of spins. The KIPA is a weakly non-linear micro-resonator engineered from the high kinetic inductance superconductor niobium titanium nitride (NbTiN), which is a material known to produce high-$Q$ and magnetic field-resilient resonators \cite{Samkharadze2016,Kroll2019}. The weak non-linearity of the KIPA allows it to act as both the ESR cavity and first-stage amplifier, where spin echo signals are amplified in-situ via three wave mixing with an applied pump tone. Compared to using a low-noise cryogenic transistor amplifier as the first stage amplifier, we demonstrate an enhancement to the signal to noise ratio of 7.5 at a magnetic field of 6.8~mT and 3.8 at 250~mT, corresponding to a more than order of magnitude reduction in measurement times.

\section{Results}

\begin{figure}[ht!]
\includegraphics[width=0.45\textwidth]{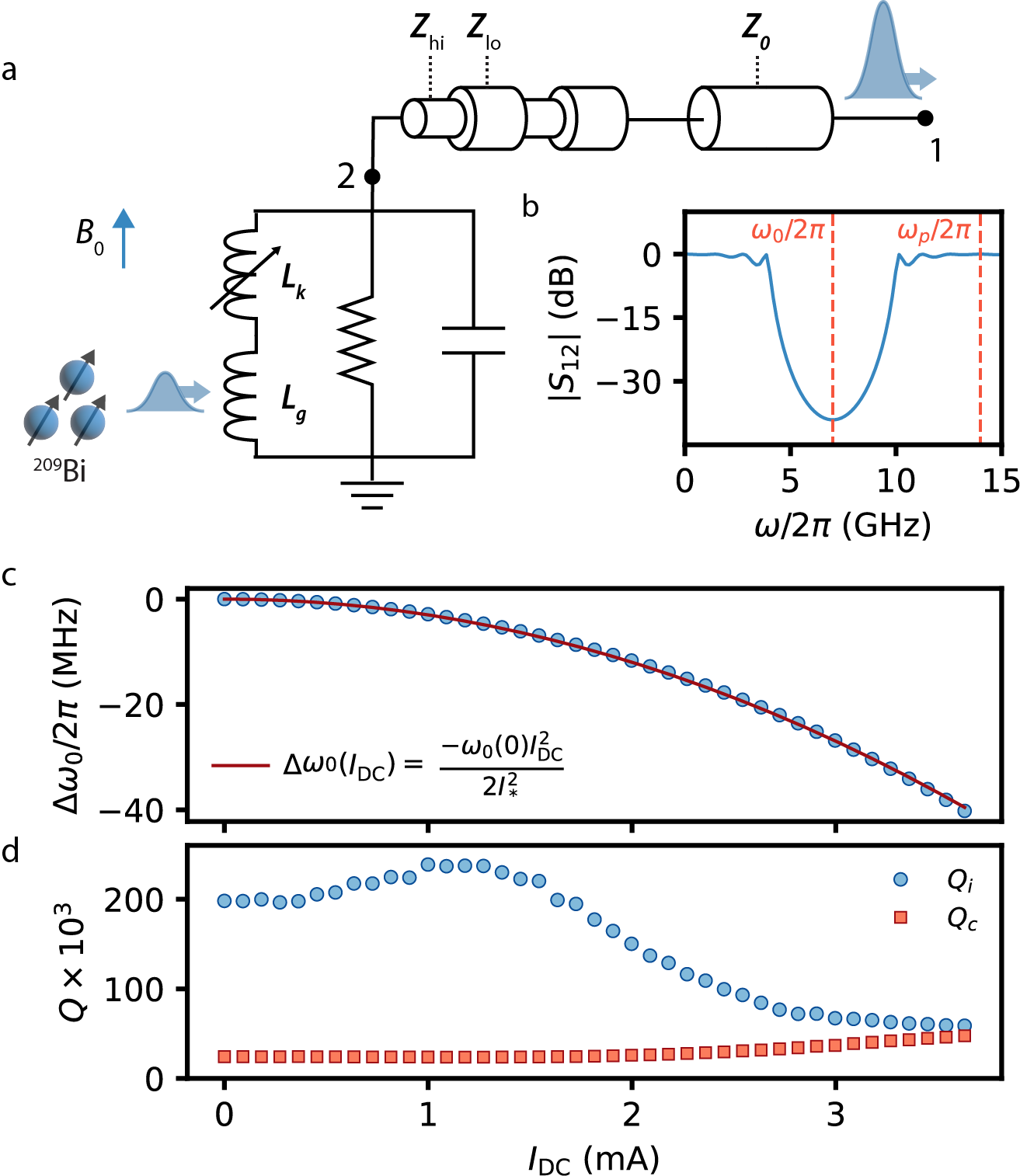}
\caption{Device design and resonator characterization. (a) A schematic for the device. The resonator is depicted as a lumped-element RLC resonator with a geometric inductance ($L_g$) that couples to an ensemble of \Bi\ and a non-linear $L_k$ which we exploit for amplification. The resonant mode is confined using a SIF, which we depict as a series of waveguides with alternating $Z_\mathrm{hi}$ and $Z_\mathrm{lo}$. (b) The frequency-dependent transmission of the SIF calculated from ABCD matrices. The frequency of the resonator and pump tones are indicated. (c,d) Parameters of the device extracted from measurements of $S_{11}$ made as a function of $I_\mathrm{DC}$. The solid line in panel c is a fit to the equation in the inset.}
\label{fig1}
\end{figure}

\subsection{Device Design and Characterization}

The device utilized in this work is based on a Kinetic Inductance Parametric Amplifier \cite{Parker2021} and consists of a $\lambda/4$ Coplanar Waveguide (CPW) resonator positioned at the end of a Stepped Impedance Filter (SIF), as depicted in Fig.~\ref{fig1}a. The SIF is constructed from a series of eight $\lambda/4$ impedance transformers with alternating high- ($Z_\mathrm{hi}$) and low-impedance ($Z_\mathrm{lo}$) (see Supplementary Materials for detailed design specifications). The filter is analogous to an optical Bragg mirror and has been used to create Purcell filters for superconducting qubits \cite{Bronn2015} and 1D-photonic crystals \cite{Liu2017}. We use the SIF to confine the resonator's mode while maintaining a galvanic connection to it; this enables the resonance frequency ($\omega_0$) to be tuned by a DC-current ($I_\mathrm{DC}$) \cite{Annunziata2010,Asfaw2017,Sigillito2017a} and facilitates amplification via three wave mixing (3WM) when a pump with frequency $\omega_p\approx 2\omega_0$ is introduced \cite{Vissers2016,Parker2021}. The predicted frequency-dependent transmission of a SIF can be calculated from ABCD matrices \cite{Pozar2012} and is plotted according to our design in Fig.~\ref{fig1}b; it shows a deep stopband centered on $\omega_0$ which serves to isolate the resonant mode from the measurement port. Notably, the filter has passbands at DC and at $\omega_p$, so that both $I_\mathrm{DC}$ and the pump can be efficiently passed to the resonator.

The device is fabricated from a 50~nm thick film of NbTiN on an isotopically-enriched {\Sitwoeight} substrate that has been ion implanted with bismuth donors at a concentration of $1\times10^{17}~\textrm{cm}^{-3}$ over a depth of approximately $1.25~\mu m$. The width of the resonator (i.e. the final $\lambda/4$ segment) center conductor is $1~\mu$m with $10~\mu$m gaps to ground and is designed to have a fundamental resonance at $\omega_0/2\pi = 7.3~$GHz with an impedance of $Z_r(\omega_0) = 240~\Omega$. The film exhibits a large kinetic inductance due to the inertia of the Cooper pairs, which displays a weak non-linear dependence on the total current $I$ according to \cite{Zmuidzinas2012}:
\begin{align}\label{eq:Lk}
    L_k(I) = L_{k0}(1+I^2/I_*^2)
\end{align}
where $L_{k0} = 3.5~\textrm{pH}/\square$ for our NbTiN film (see Supplementary Materials for details) and $I_*$ is a constant related to the critical current \cite{Parker2021}. This non-linearity allows the resonant frequency to be tuned through the application of a DC current $I_\textrm{DC}$ and facilitates amplification. The spins couple to the magnetic field produced by the device, or equivalently its geometric inductance $L_g$. It is therefore important to balance the amount of kinetic and geometric inductance present to ensure a sufficient non-linearity for performing amplification without substantially reducing the coupling to the spins (a detailed discussion on this is presented in the Supplementary Materials). 

We use a Vector Network Analyzer (VNA) to measure the response of the device in reflection ($S_{11}$) without amplification and determine the experimental value of $\omega_0/2\pi$ to be 7.233~GHz at $I_\mathrm{DC}=0~$mA. We also determine the internal quality factor ($Q_i$) and coupling quality factor ($Q_c$) by fitting the combined amplitude and phase response \cite{Probst2015} and display the $I_\mathrm{DC}$ dependence of the fit results in Figs.~\ref{fig1}c-d. The device can be tuned over a range of 40~MHz by applying up to $I_\mathrm{DC}=3.63~$mA, before the superconducting film transitions to the normal state. We expect a quadratic dependence of the change in resonance frequency with the applied DC current $\Delta\omega_0(I_\mathrm{DC}) = -\omega_0(0)I_\mathrm{DC}^2/(2I_*^2)$, which allows us to extract $I_*=34.5~$mA for this device. $Q_i$ decreases monotonically for $I_\mathrm{DC}>1.3~$mA and $Q_c$ grows by a factor of two; the combined effect is that the device approaches critical coupling ($Q_i=Q_c$) as $I_\mathrm{DC}$ is raised.

\begin{figure*}[t!]
\includegraphics{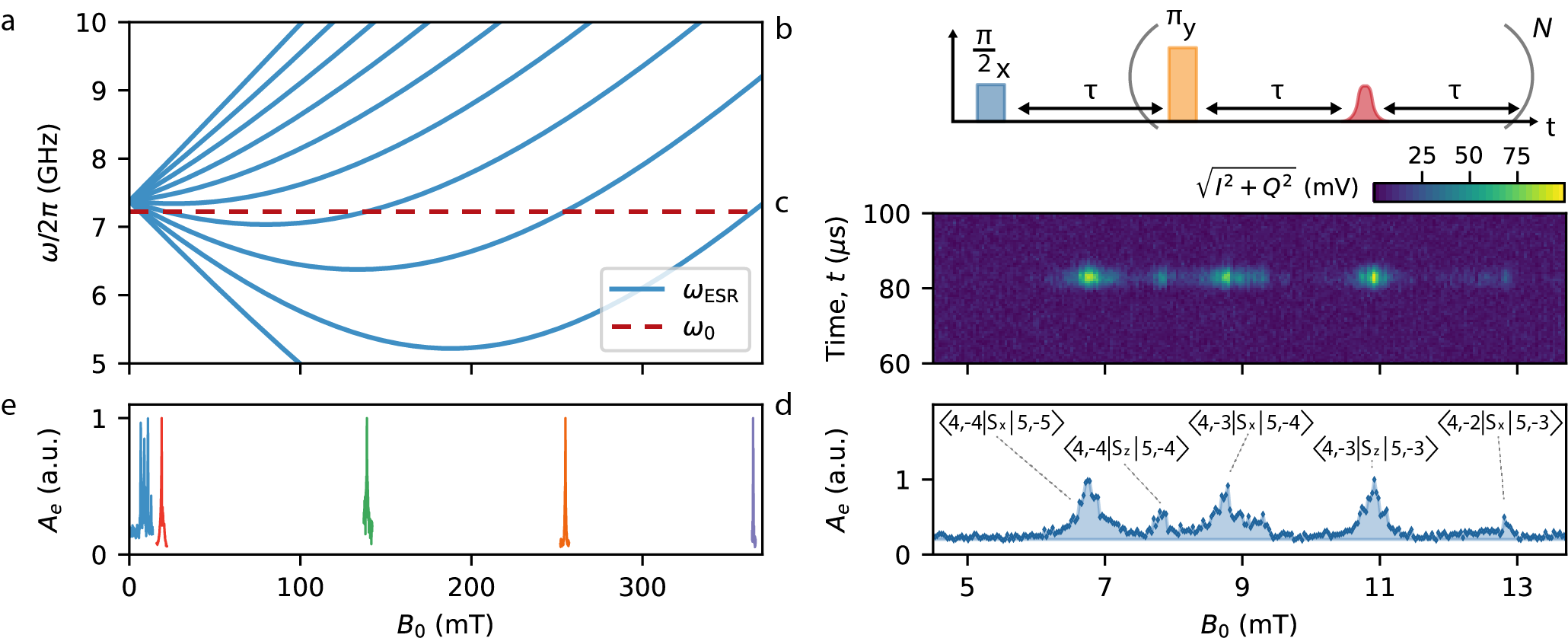}
\caption{ESR measurements of \Bi\ donors in Si using a KIPA. (a) The allowed ESR transition frequencies for \Bi\ in Si as a function of $B_0$. We can measure ESR with the KIPA when $\omega_\mathrm{ESR}=\omega_0$. (b) The CPMG sequence applied to detect the spins. We use $\tau=75~\mu$s and $N=200$, averaging the echo produced by each of the $N$ refocusing pulses to increase the SNR. (c) The homodyne-demodulated signal in the time-domain as a function of $B_0$, measured with the CPMG sequence shown in panel b. The bright features correspond to a spin echo signal. (d) The integrated spin echo signal from panel c. We label the five peaks according the ESR transitions we expect from calculations of the spin Hamiltonian. (e) Measurements of the $S_x$ transitions between 0~mT and 370~mT. Each measurement is independently normalized.}
\label{fig2}
\end{figure*}

\subsection{Pulsed ESR Measurements}
Our measurements are performed at a temperature of $400~$mK, where the \Bi\ donors bind one additional valence electron compared to the surrounding Si atoms. The donor-bound electron and its nucleus are coupled via the contact hyperfine interaction, $H_A/\hbar = A \bf{S}\cdot \bf{I}$, where $A/2\pi = 1.478~$GHz \cite{Mohammady2010} and $\bf{S}$ ($\bf{I}$) represents the electron (nuclear) spin operator. At fields $B_0 < 100$~mT, the contact hyperfine interaction is of comparable strength to the Zeeman interaction, $H_{B}/\hbar = B_0(\gamma_e S_z + \gamma_n I_z)$, where $\gamma_e/2\pi=27.997~$GHz/T and $\gamma_n/2\pi=-6.96~$MHz/T. Here the eigenstates of the spin Hamiltonian $H = H_A + H_B$ are best described by the total spin $\bf{F} = S + I$ and its projection onto $B_0$, $m_F$. The twenty electron-nuclear states $\ket{F,m_F}$ are divided into an upper manifold of eleven states with $F=5$ and a lower manifold of nine states with $F=4$. The ESR-allowed transitions, which are driven through the $S_x$ spin operator, are calculated and displayed in Fig.~\ref{fig2}a, where we plot the frequency of the transitions as a function of the magnetic field.


The static magnetic field $B_0$ is applied in the plane of the device and parallel to the long-axis of the resonator. This alignment is chosen so that the magnetic field $B_1$ produced by the $\lambda/4$ resonator which drives spin resonance is perpendicular to $B_0$ for the spins located underneath the resonator, in order to probe the $S_x$ ESR transitions. To perform ESR, we tune $B_0$ so that $\omega_\mathrm{ESR}=\omega_0$ (horizontal dashed line in Fig.~\ref{fig2}a).

We detect the spins using a Carr-Purcell-Meiboom-Gill (CPMG) pulse sequence (Fig.~\ref{fig2}b). Due to the long spin coherence times of \Bi\ donors in isotopically enriched {\Sitwoeight} \cite{Wolfowicz2013}, we are able to repeatedly refocus the ensemble, collecting a spin echo each time we do so. We use $N=200$ refocusing pulses and average the echo signals to increase the signal-to-noise-ratio (SNR) of our measurement \cite{Bienfait2016}. The result over a small field range is shown in Fig.~\ref{fig2}c, where we plot the amplitude of the homodyne-demodulated signal, $\sqrt{I(t)^2 + Q(t)^2}$, as a function of $B_0$. In Fig.~\ref{fig2}d we also plot the integrated echo signal $A_e =(1/T_e) \int_{0}^{T_e} \sqrt{I(t)^2 + Q(t)^2} dt$, where $T_e$ is the duration of the spin echo signal. By comparing the measured spectrum to an exact diagonalization of the Hamiltonian $H$, we can identify the transitions present in this field sweep. Three of the transitions are coupled by the $S_x$ operator. These are the typical ESR transitions that were expected for the orientation of the resonator and the external magnetic field. Two of the transitions, however, are coupled by the $S_z$ spin operator. In the $\ket{F,m_F}$ basis these correspond to transitions of the type $\ket{4,m_F} \leftrightarrow \ket{5,m_F}$ and are driven by a longitudinal magnetic field (i.e. when $B_1$ is parallel to $B_0$). Though they are forbidden at high field, these transitions can be observed at low field due to the hyperfine coupling, $H_A$ \cite{Pla2018a}. This suggests that the resonant mode of our device is not fully confined to the last $\lambda/4$ section. Finite element simulations confirm this (see Supplementary Materials) and show that at $\omega_0$ there is an appreciable $B_1$ field in the last $Z_\mathrm{hi}$ segment of the SIF, which is oriented perpendicular to the resonator (see Supplementary Materials for a full device schematic) and where $B_1 \parallel B_0$ for spins located underneath the inner conductor of the CPW.

In addition to the five transitions measured at $B_0<13~$mT, we demonstrate the ability to measure each $S_x$ transition out to $370~$mT (Fig.~\ref{fig2}e). Aside from the use of NbTiN as the superconducting material and the in-plane alignment of $B_0$, we did not implement any specific features to enhance our device's compatibility with magnetic fields, such as the inclusion of vortex pinning sites \cite{Kroll2019}. By converting the magnetic field to the equivalent ESR frequency of a free electron, we see that this range of fields encompasses the X-band frequency range ($8-10$~GHz) common for commercial ESR spectrometers.

\subsection{In-Situ Amplification of Spin Echoes}

\begin{figure}[t!]
\includegraphics[width=0.45\textwidth]{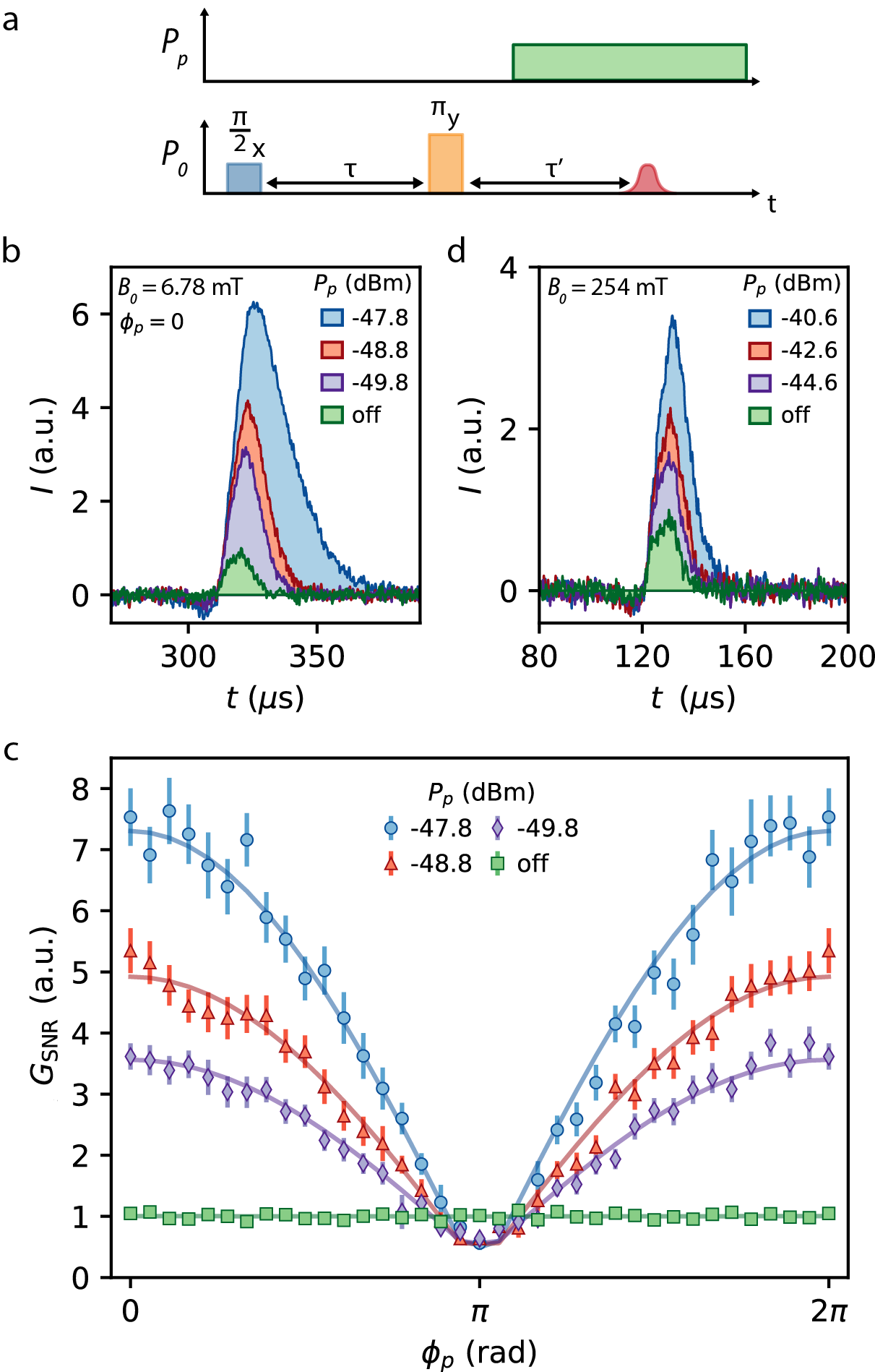}
\caption{Degenerate amplification of spin echoes. (a) A modified Hahn echo pulse sequence where a strong parametric pump at frequency $\omega_p=2\omega_0$ and power $P_p$ is supplied following the refocusing pulse. The device functions as a typical high-$Q$ resonator for the first half of the pulse sequence and as a degenerate parametric amplifier during the period the spins induce a signal in the device. (b) Amplified spin echoes measured along the $I$-quadrature for several $P_p$. For these measurements $\phi_p=0$, $I_\mathrm{DC}=3.0~$mA, and $B_0=6.78~$mT. The data is normalized to the measurement with the pump off. (c) $G_\mathrm{SNR}$ measured at the same set point as in panel b. The improvement to the SNR is $\phi_p$-dependent because the amplifier is operated in degenerate mode. The errorbars correspond to the standard error of the mean and the solid lines are guides to the eye. (d) Amplified spin echoes measured with $I_\mathrm{DC}=2.0~$mA and $B_0=254~$mT. Note that at this setpoint we do not control for $\phi_p$. The data is normalized to the measurement with the pump off.}
\label{fig3}
\end{figure}

We now seek to use the weak non-linearity of the NbTiN film to perform in-situ amplification of the spin echo signals. A pump tone is applied to the device at a frequency of $\omega_p$ and amplification occurs as energy transfers from the pump to the spin echo signal at $\omega_0$. The quadratic dependence of the kinetic inductance with current (see Eq.~\ref{eq:Lk}) naturally lends itself to a four wave mixing (4WM) process, where energy conservation requires $2\omega_p = \omega_0+\omega_i$, with $\omega_i$ corresponding to the `idler' frequency -- a tone produced during amplification. The application of a DC current lowers the order of the non-linearity \cite{Malnou2021,Vissers2016,Parker2021} and enables three wave mixing processes, where energy conservation dictates $\omega_p = \omega_0+\omega_i$. 3WM is preferable in amplification as it provides a large spectral separation between the pump and signal, allowing the pump to be filtered out of the detection chain. In addition, for kinetic inductance amplifiers 3WM is a more efficient process than 4WM, requiring lower pump powers for equivalent gains \cite{Vissers2016}. 
To amplify the spin echoes in-situ via 3WM, we first DC current bias the device with $I_\mathrm{DC}=3.0~$mA. Next, we apply a standard Hahn echo pulse sequence with the addition of a pump tone at the frequency $\omega_p=2\omega_0$ and power $P_p$, which is sent $50~\mu$s following the trailing edge of the refocusing pulse (Fig.~\ref{fig3}a). The device therefore functions as a resonator during the delivery of the spin control pulses and as a resonant parametric amplifier when the spin echoes are detected. For a pump frequency precisely twice the signal frequency (as used here), the signal and idler tones become degenerate and the gain depends on the relative phase between the signal and pump \cite{Parker2021}. In Fig.~\ref{fig3}b we compare spin echoes measured at $B_0 = 6.78~$mT on the $\bra{4,-4}S_x\ket{5,-5}$ transition. The echoes are aligned along the $I$ signal quadrature and presented for several different $P_p$. The pump phase $\phi_p$ is tuned at each $P_p$ to maximize the gain and therefore spin echo amplitude. For $P_p = -47.8~$dBm, the maximum echo amplitude is greater than $7$ times larger than that of an equivalent measurement with the pump turned off, where the first stage amplifier corresponds to a cryogenic low noise high electron mobility transistor (HEMT) amplifier (see Supplementary Materials for setup details). Interestingly, the duration of the echo also lengthens when $P_p$ is increased. We attribute this to the finite Gain-Bandwidth Product (GBP) of the KIPA that is common to many resonant parametric amplifiers, which in our measurement extends the length of time it takes for the amplified intracavity field to decay.

\subsection{Signal to Noise Ratio}
We define the amplitude SNR of the Hahn echo measurements as:

\begin{equation}
\mathrm{SNR} = \frac{\left.\frac{1}{T_e}\int_{0}^{T_e} I(t) dt\right|_e}{\left.\sqrt{\frac{1}{T_e}\int_{0}^{T_e} I^2(t) dt}\right|_b}
,\end{equation}

\noindent where the $e$ and $b$ subscripts refer the the experimental pulse sequence (which produces a spin echo) and a blank pulse sequence (which gives a measure of the noise), respectively. For the blank sequence, we omit the $\pi_y$ refocusing pulse from the Hahn echo sequence so that no spin echo is produced. The amplitude SNR is therefore the ratio of the mean amplitude of the spin echo and the root-mean-square (RMS) of the noise.

In our experiments the duration of the spin echo depends on $P_p$, so to calculate the SNR we keep the window fixed to the duration of the shortest echo, corresponding to the measurement with the pump off (e.g. $T_e \in [310-330]~\mu$s in Fig.~\ref{fig3}b). This ensures that the bandwidth of the noise is equal when we compare the SNR for measurements taken with different $P_p$. The improvement to the amplitude SNR when parametrically pumping the device is then given by:

\begin{equation}
G_\mathrm{SNR} = \frac{\mathrm{SNR}|_\text{pump on}}{\mathrm{SNR}|_\text{pump off}}
.\end{equation}

In Fig.~\ref{fig3}c we demonstrate that $G_\mathrm{SNR}$ is phase dependent, displaying a period of $2\pi$ with the pump phase $\phi_p$. This is evidence that the amplifier indeed acts in degenerate mode. For these measurements we align the maximum amplitudes of the different traces at $\phi_p=0$, where we find $G_\mathrm{SNR}(\phi_p=0)= 7.5 \pm 0.5$ at the highest pump power $P_p=-47.8~$dBm. At this power we also measure $G_\mathrm{SNR}(\phi_p=\pi)= 0.6 \pm 0.1~$, demonstrating that the spin echo signal is deamplified for certain phases.

In Fig.~\ref{fig3}d we show similar measurements taken at $B_0 = 254~$mT, corresponding to the $\bra{4,-3}S_x\ket{5,-2}$ spin transition. Due to a gradual decline of $Q_i$ as we increase $B_0$, we choose to work with $I_\mathrm{DC}=2.0~$mA where $\omega_0/2\pi=7.218~$GHz, $Q_i=37.9\times 10^3$ and $Q_c=27.3 \times 10^3$. At this field we manage to enhance the amplitude of the spin echo signal by up to a factor of $3.4$ relative to a measurement without the pump applied, corresponding to $G_\mathrm{SNR}=3.8\pm 0.3$. Notably, this enhancement is achieved without explicitly controlling for $\phi_p$. The SNR of the spin signal at this transition was smaller than at 6.78~mT, requiring us to increase the number of measurement averages. Due to the limited hold time of our pumped \Hethree\ cryostat, this prevented us from measuring $G_\mathrm{SNR}$ as a function of $\phi_p$. Instead, we kept the microwave sources phase locked but randomized $\phi_p$ between repetitions, such that the $G_\mathrm{SNR}$ we report is effectively the average of $G_\mathrm{SNR}(\phi_p)$. This SNR enhancement could therefore be improved by a factor of 2 with appropriate choice of pump phase. 

\begin{figure}[t!]
\includegraphics[width=0.45\textwidth]{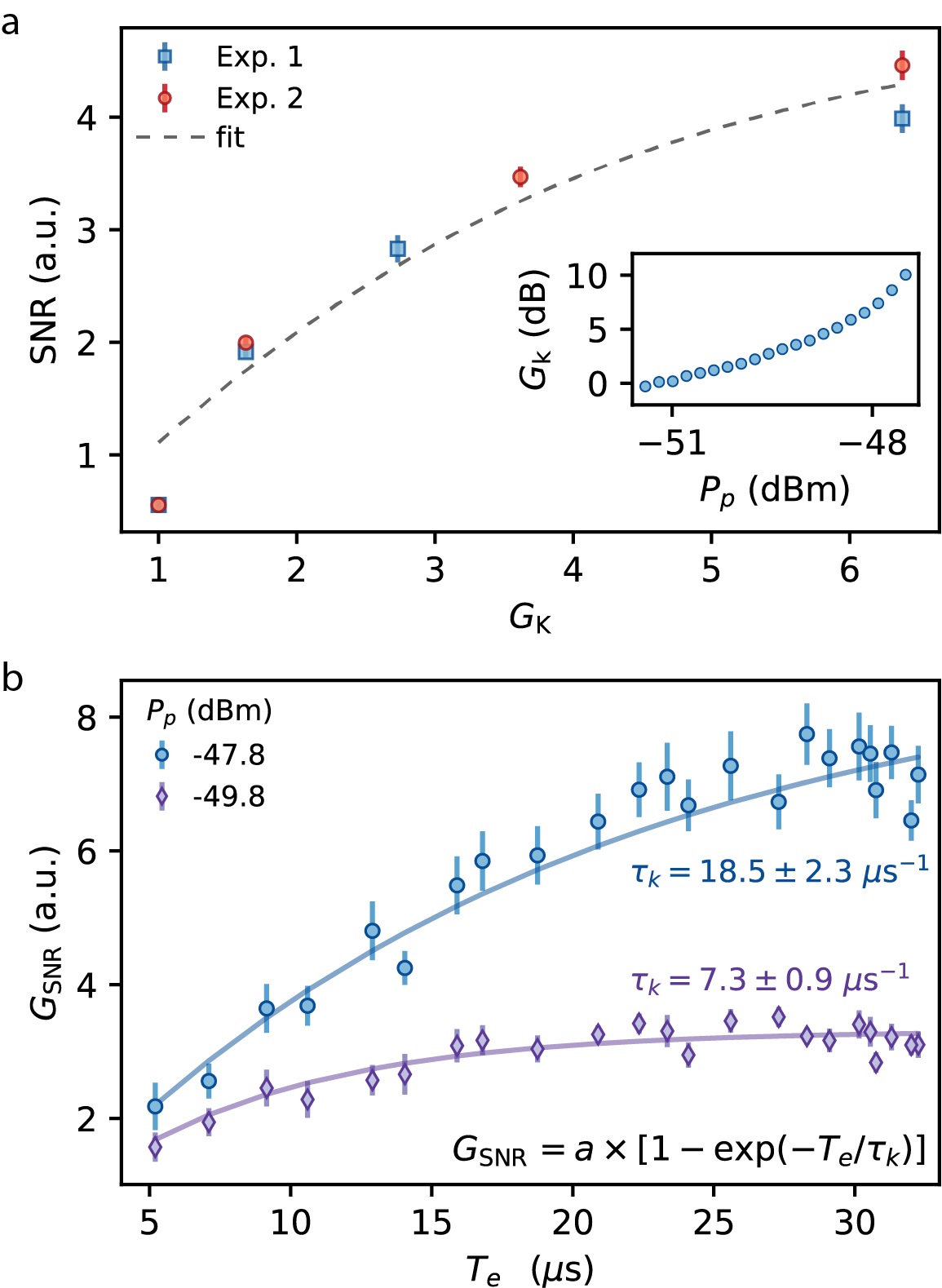}
\caption{SNR gain and amplifier bandwidth. (a) $\mathrm{SNR}$ measured as a function of $G_k$. The dashed line is a fit of the data to a model derived from cavity input-output theory (see Supplementary Materials). Experiments one and two are equivalent experiments performed on different days. The data for experiment two was scaled by a factor of 1.33 so that the SNR with the pump off matches experiment one. Inset: $G_k$ measured as a function of $P_p$. (b) $G_\mathrm{SNR}$ measured as a function of $T_e$. The solid lines are fits to the equation in the inset.}
\label{fig4}
\end{figure}

Next we investigate the dependence of $G_\mathrm{SNR}$ on the amplitude gain of the amplifier ($G_k$, Fig.~\ref{fig4}a). $G_k$ is measured using a spectrum analyzer to assess the degenerate gain of a coherent signal with frequency $\omega_0$ reflected off the input of the KIPA and with $\phi_p$ chosen to maximize the gain (inset of Fig.~\ref{fig4}a). We see that $G_\mathrm{SNR}$ initially grows rapidly with $G_k$, but begins to saturate at high $G_k$. This can be explained by considering the three contributions to the total noise: noise on the spin echo signal itself ($n_s$), e.g. vacuum, thermal and spontaneous emission noise, noise added by the KIPA ($n_k$), and noise added by the components following the KIPA ($n_\mathrm{sys}$). As we show using a model derived from cavity input-output theory in the Supplementary Materials, the SNR will grow with $G_k$ so long as $G_k^2(n_s + n_k) \lesssim n_\mathrm{sys}$. In our measurements, $n_\mathrm{sys}$ is dominated by insertion loss and the noise added by the cryogenic HEMT amplifier. The fit shown in Fig.~\ref{fig4}a is based on our model and allows us to estimate the ratio of the system noise to the noise on the spin echo signal and the parametric amplifier added noise (see Supplementary Materials). We find that $n_\mathrm{sys}/(n_s+n_k) \approx 20$, which agrees with our estimates for the system noise $n_\mathrm{sys} \gtrsim 12$~photons per quadrature and for $(n_s+n_k) \approx 0.6$~photons per quadrature (see Supplementary Materials). We note that we were not able to reach the high-gain limit in the present experiments. This occurs when $G_k$ is large enough such that $G_k^2(n_s+n_k)\gg n_\mathrm{sys}$, whereafter $G_\mathrm{SNR}$ becomes independent of $G_k$. When raising $G_k$ beyond $\sim 6.5$ we observed a hysteretic onset of parametric self-oscillations, which is the focus of a future study.

To explore the effects of the constant GBP of the KIPA, we perform an experiment where we amplify the spin echoes while varying $T_e$, which we achieve by varying the duration of the tipping and refocusing pulses in a Hahn echo sequence. For long $T_e$ the bandwidth of the spin signal is smaller than the bandwidth of the KIPA and the entire echo signal is amplified. However, as $T_e$ is made shorter, the bandwidth of the spin signal can exceed the amplification bandwidth of the device, which reduces $G_\mathrm{SNR}$. Since the device has a fixed GBP we expect higher gains to produce smaller bandwidths. In Fig.~\ref{fig4}b we compare $G_\mathrm{SNR}(T_e)$ measured for two different $P_p$ and find both experiments are well-described by the function $G_\mathrm{SNR} = a[1-\exp(-T_e/\tau_k)]$ where $a$ and $\tau_k$ are constants. From the fits we find $a_1/a_2= 2.7 \pm 0.2$ and $\tau_{k1}/\tau_{k2} = 2.5 \pm 0.4$, where the subscript $1(2)$ corresponds to the measurement with $P_p=-47.8~$dBm ($-49.8$~dBm). The close agreement of these two ratios confirms that the GBP is constant for the two experiments, since an increase in echo amplitude is associated with a corresponding increase in the echo duration required to saturate the SNR gain.

\subsection{Spin Sensitivity}
The sensitivity of ESR measurements is often reported as the minimum number of spins ($N_\mathrm{min}$) required to achieve an SNR of unity for the detection of a single spin echo. By combining finite element modelling of the electromagnetic field distributions in the KIPA with the bismuth ion-implantation profile, we estimate the total number of donors within the magnetic field of the resonant mode of the KIPA to be approximately $6 \times 10^6$ (see Supplementary Materials). Most of these donors do not contribute to the spin echo signals as they (i) don't populate the spin transitions probed, or (ii) are not excited by the selective pulses used in the Hahn echo sequence, or (iii) they do not rotate by the correct angles in the pulse sequences as a result of the inhomogeneous coupling of the spin ensemble to the resonator. See the Supplementary Materials for a detailed discussion and numerical estimates for each of these factors. Accounting for these effects, we estimate that the number of spins that contribute to the spin echo signals in our measurements is approximately $1\times10^4$. Using the SNRs measured in Fig.~\ref{fig3}, we find $N_\mathrm{min}\approx 2\times 10^4$ when the pump is off, which improves up to $N_\mathrm{min} \approx 2.8\times 10^3$ by performing in-situ amplification of the spin echoes. 

\section{Discussion}
The enhancement observed in the spin sensitivity and SNR when amplifying with the KIPA can be attributed to an approximate 20 times reduction in the system noise temperature, which results from using a low-noise parametric amplifier and eliminating the insertion loss between our micro-resonator and amplifier. The success of this approach is evident when one compares the maximum $G_\mathrm{SNR}=7.5\pm 0.5$ to previous studies using JPAs separated from the micro-resonator and measured at 20~mK ($G_\mathrm{SNR}=11.7$ \cite{Bienfait2016a} and $G_\mathrm{SNR}=5.9$ \cite{Eichler2017}); we achieve a similar improvement to the SNR in our measurements despite the fact that we operate at a higher temperature (400~mK) where the noise exceeds the quantum limit.

There are several avenues for improving the SNR achieved here. Cooling the device down below 100~mK would reduce both the thermal noise on the signal and the excess noise added by the KIPA. In addition, the low temperature would increase the spin polarization, which is $p = \tanh{[\hbar\omega_0/(2k_BT)]}=0.40$ in our current measurements, leading to an an overall improvement to the SNR by at least a factor of five. The non-linearity of the NbTiN film could also be optimized to maximize the measurement SNR. Decreasing the amount of kinetic inductance relative to the geometric inductance would enhance the spin to resonator coupling strength and the total number of spins that contribute to the echo signals, at the expense of requiring larger pump powers. We estimate that reducing the kinetic inductance fraction $L_k/(L_k+L_g)$ from 0.8 in the present device to 0.4 would boost the SNR by a factor of 2-3, and require only a modest increase ($< 5$~dB) in pump power to maintain the gains achieved here (see Supplementary Materials for further discussion).

Owing to the large critical temperature of NbTiN ($T_c \sim 13$~K), the KIPA can readily be operated at higher temperatures ($\sim 2$~K), such as those accessible in pumped Helium-4 X-band ESR spectrometers. Indeed, NbTiN kinetic inductance amplifiers have recently been shown to operate at temperatures up to 4~K \cite{malnou2022}. Combining high temperature operation with the high magnetic field compatibility demonstrated here, this device could translate the high-sensitivities achieved in bespoke quantum-limited ESR spectrometers to more conventional ESR operating conditions and systems. Along with recent studies of four-wave mixing parametric amplifiers made from NbN \cite{Xu2022} and NbTiN \cite{Khalifa2022}, our demonstration here of a magnetic field-resilient superconducting parametric amplifier is not only promising for measurements of spin ensembles but a broad class of quantum experiments that require magnetic fields \cite{Clerk2020}.

\section{Materials and Methods}

\paragraph{Device and measurement setup} The device is fabricated from a $50~$nm film of NbTiN on isotopically enriched \Sitwoeight\ (500~ppm {\Sitwonine}). The {\Bi}-donors are implanted uniformly across the entire substrate with a concentration of $10^{17}~\text{cm}^{-3}$ over a depth of $1.25~\mu$m. All measurements were performed at $400$~mK using a pumped \Hethree\ cryostat and a homemade spectrometer. Further details on the device design and measurement setup are provided in the Supplementary Materials.

\paragraph{Data processing}

The measurements shown in Figs.~\ref{fig3}b,c were collected over an $18$~hr period (approaching the maximum hold time for the pumped \Hethree\ cryostat). For each repetition of the experiment the settings of $P_p$ and $\phi_p$ were selected in pseudo-random order, with the control experiment for each setting performed immediately after each measurement. For each repetition we recorded the average of thirty shots for each setting of $P_p$, $\phi_p$, and their corresponding control sequences. A total of seven repetitions was completed so that the data shown in Fig.~\ref{fig3}b,c corresponds to 210 shots total per data point. The pulse sequences were executed with a 1~Hz frequency, which is the same order of magnitude as $1/T_1$ (see Supplementary Materials). The experiments collected in Fig.~\ref{fig3}d,e are comprised of a total of 5000 shots of the pulse sequence, where $\phi_p$ was randomized for each measurement.

The SNR and $G_\mathrm{SNR}$ are calculated from post-processed data. From the raw data, we subtract a constant offset from both the $I$ and $Q$ quadratures, originating from components in the detection chain (amplifiers, mixers, digitizer etc). We then down-sample the data from the native $2~$ns resolution of the digitizer to $50~$ns and use a $1~$MHz digital low-pass filter to reduce the noise. Next, we rotate the data in the $IQ$-plane such that the echo is aligned along the $I$-quadrature (by minimizing $\int_{t_1}^{t_2} |Q(t)| dt$). For the phase-sensitive experiments we correct for phase drift by fitting the normalized integrals $A_I(\phi_p)$ with the phenomenological function $|\sin(\phi_p - \pi)|$. We note that the phase offsets acquired from the fits of the measurements with spin echoes are also applied to the corresponding control experiments to ensure that any imbalance in the gain of the $I$ and $Q$ detection channels is accounted for. The data shown in Fig.~\ref{fig3}c corresponds to the mean and standard error of the mean of the seven repetitions. 

The data shown in Fig.~\ref{fig4}a is from the same experiment as Fig.~\ref{fig3}b,c where $\phi_p=0$, and an equivalent measurement taken on another day. The SNR reported in Fig.~\ref{fig4}a has been re-scaled to that of a single shot as $\mathrm{SNR}=\overline{\mathrm{SNR}}/\sqrt{M}$, where $\overline{\mathrm{SNR}}$ is the SNR found from the mean of $M$ shots of the Hahn echo pulse sequence. Experiment 1 (Experiment 2) used $M = 30$ ($M = 20$) shots. The SNR for Experiment 2 was found to be smaller than for Experiment 1, which is likely due to a small drift in $B_0$ or the resonator's frequency. To account for this, we scale the data in Fig.~\ref{fig4}a by a factor of 1.33 so that the SNR of Experiment 2 with the pump off matches the SNR of Experiment 1 with the pump off. The unscaled-SNR and corresponding $N_\mathrm{min}$ for both experiments are summarized in the Supplementary Materials.

To extend $T_e$ for the measurements shown in Fig.~\ref{fig4}b we varied the duration of the $\pi/2$ and $\pi$ pulses in a Hahn echo sequence between $2~\mu$s and $25~\mu$s. $T_e$ was then found using a threshold applied to the mean of the spin signal measured with the pump off.

\section{Acknowledgments}
J.J.P. acknowledges support from an Australian Research Council Discovery Early Career Research Award (DE190101397). J.J.P. and A.M. acknowledge support from the Australian Research Council Discovery Program (DP210103769). A.M. is supported by the Australian Department of Industry, Innovation and Science (Grant No. AUS-MURI000002). T.S. was supported by the Office of Fusion Energy Sciences, U.S. Department of Energy, under Contract No. DE-AC02-05CH11231. W.V. and J.S.S. acknowledge financial support from Sydney Quantum Academy, Sydney, NSW, Australia. This research has been supported by an Australian Government Research Training Program (RTP) Scholarship. The authors acknowledge support from the NSW Node of the Australian National Fabrication Facility. We acknowledge the AFAiiR node of the NCRIS Heavy Ion Capability for access to ion-implantation facilities at EME, ANU. We thank Robin Cantor and STAR Cryoelectronics for sputtering the NbTiN film.

\section{Author contributions}
W.V. and M.S. performed the experiments. W.V. analyzed the data. M.S. and J.J.P. designed the device. M.S. fabricated the device. J.S.S. performed device simulations. T.S. provided the isotopically enriched silicon substrate and B.C.J. and J.C.M. performed the \Bi\ implantation. D.P. built the measurement setup. J.J.P. and A.M. supervised the project. W.V. and J.J.P. wrote the manuscript with input from all authors.   

\section{Additional information}
The authors declare no competing financial interests. Online supplementary information accompanies this paper. Correspondence and requests for materials should be addressed to J.J.P.

\bibliographystyle{unsrt}
\bibliography{Papers-spin_amp_refs}

\end{document}


\title{Supplementary Materials: In-situ amplification of spin echoes within a kinetic inductance parametric amplifier}

\author{Wyatt Vine}
\thanks{These two authors contributed equally}
\affiliation{School of Electrical Engineering and Telecommunications, UNSW Sydney, Sydney, NSW 2052, Australia}
\author{Mykhailo Savytskyi}
\thanks{These two authors contributed equally}
\altaffiliation{Present address: IQM Finland Oy, Espoo 02150, Finland}
\affiliation{School of Electrical Engineering and Telecommunications, UNSW Sydney, Sydney, NSW 2052, Australia}
\author{Daniel Parker}
\affiliation{School of Electrical Engineering and Telecommunications, UNSW Sydney, Sydney, NSW 2052, Australia}
\author{James Slack-Smith}
\affiliation{School of Electrical Engineering and Telecommunications, UNSW Sydney, Sydney, NSW 2052, Australia}
\author{Thomas Schenkel}
\affiliation{Accelerator Technology and Applied Physics Division, Lawrence Berkeley National Laboratory, Berkeley, California 94720, USA}
\author{Jeffrey C. McCallum}
\affiliation{School  of  Physics,  University  of  Melbourne,  Melbourne,  Victoria  3010,  Australia}
\author{Brett C. Johnson}
\affiliation{Centre of Excellence for Quantum Computation and Communication Technology, School of Engineering, RMIT University, VIC, 3001, Australia}
\author{Andrea Morello}
\affiliation{School of Electrical Engineering and Telecommunications, UNSW Sydney, Sydney, NSW 2052, Australia}
\author{Jarryd J. Pla}
\email[]{jarryd@unsw.edu.au}
\affiliation{School of Electrical Engineering and Telecommunications, UNSW Sydney, Sydney, NSW 2052, Australia}

\date{\today}

\maketitle

\tableofcontents

\section{Experimental setup}\label{sec:setup}

 \begin{figure}[ht]
 \centering
 \includegraphics[width=0.7\textwidth,keepaspectratio]{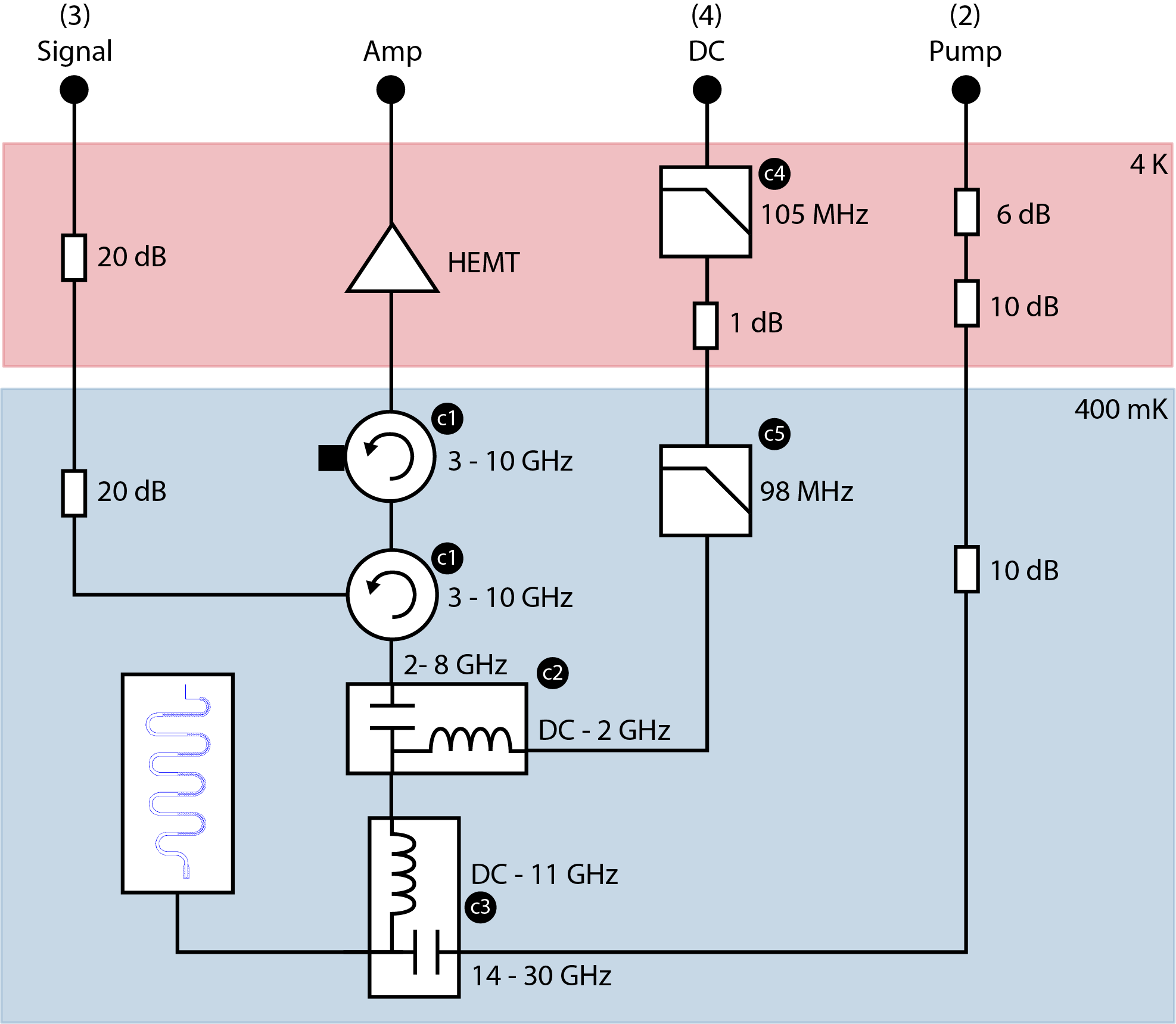}
 \caption[]{The measurement setup. The components are (c1) Raditek RADC-4-10-Cryo circulator, (c2) Marki DPXN4, (c3) Marki DPX1114, (c4) Mini-Circuits VLF-105+, (c5) Mini-Circuits SLP-100+.}
 \label{fig-sm:setup}
 \end{figure}

All measurements were performed at 400~mK using a pumped \Hethree\ cryostat. A schematic of the measurement setup is shown in Fig.~\ref{fig-sm:setup}. It consists of four semi-rigid coax cables between room temperature and the \Hethree\ stage of the cryostat. A description of each measurement line is provided below.

\begin{itemize}
\item{The signal line was used to deliver resonant power to the device ($\omega_0/2\pi \approx 7.2$~GHz). It had a total of 40~dB of fixed attenuation, split across the two stages of the cryostat. We measure the attenuation from the room temperature input to the device to be $-52\pm 1$~dB at $\omega_0$ when the system is cold (by combining $|S_{21}|$ measurements made across multiple cooldowns).}
\item{The DC line was used to supply a DC current to the device and was low-passed filtered with a cut-off of $100$~MHz.}
\item{The pump line was used to supply high frequency microwave tones at $2\omega_0$. It had a total of 26~dB of fixed attenuation. The total attenuation at $\omega_0$ is estimated to be more than $60$~dB.}
\item{The amplification line contains a single HEMT at the 4~K. It is connected to the signal line via a circulator and is isolated from the device via a second $50~\Omega$ terminated circulator. The connection between $4~$K and the amplifier is a superconducting NbTi coax cable, to reduce loss.}
\end{itemize}

The general goal of this setup is to combine signals across the three distinct frequency ranges together at $400$~mK so that the device can be measured via a single port. We accomplish this using a bias-tee to combine the DC and signal tones (Marki DPXN4) and a diplexer to further combine the pump (Marki DPX1114). In addition, each of the three lines is attenuated by at least $50~$dB at $\omega_0$, greatly surpassing the temperature difference between room temperature and the coldest stage which is $10\log_{10}(300~\text{K}/400~\text{mK}) \approx 30~$dB.

\begin{figure}[ht]
	\centering
	\includegraphics[width=0.7\textwidth]{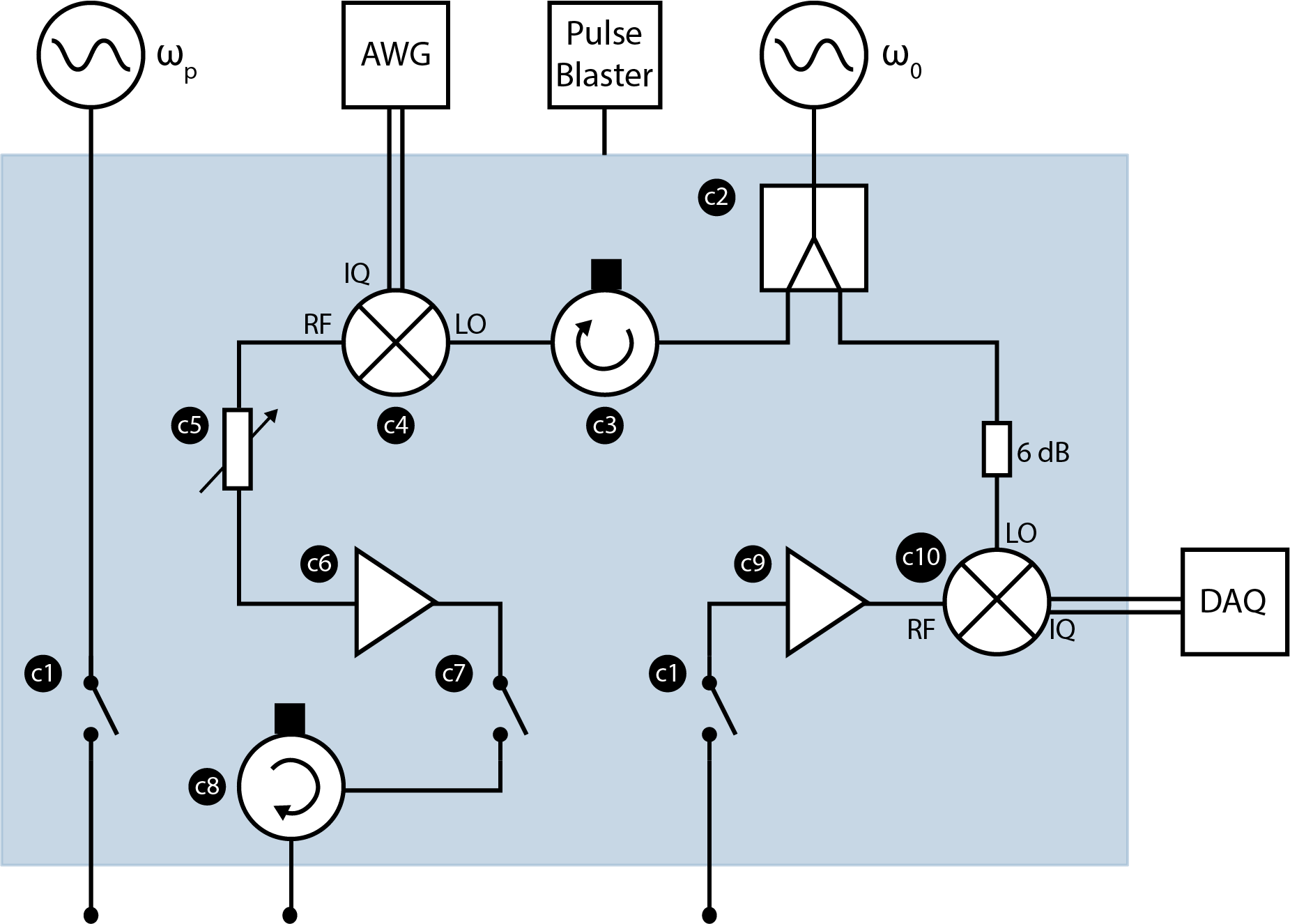}
	\caption[]{Schematic for a home-built spectrometer for pulsed ESR measurements. c1: single-pole double-throw microwave switch, c2: Marki PD-0R510, c3: Ditom D3C4080, c4: Marki IQ-4509LXP, c5: Mini-Circuits RUDAT-13G-60, c6: Mini-Circuits ZVE 3W-183+, c7: double-pole quad-throw microwave switch, c8: Ditom D3C4080, c9: Mini-Circuits ZX60-05113LN+, c10: Polyphase AD60100B.}
	\label{fig-sm:kubrick}
\end{figure}

All of the pulsed ESR measurements were obtained with a home-built spectrometer depicted in Fig.~\ref{fig-sm:kubrick}. The system has two main arms: one for generating phase-controlled pulses, and one for performing homodyne demodulation of the reflected signals. Pulses are generated from a Local Oscillator (LO) with fixed power using two channels of an Arbitrary Waveform Generator (AWG) with a microwave IQ-mixer. To extend the dynamic range of the system, a programmable attenuator can be used to vary the output power of the system. To ensure no unintended power leaks out of the box, a fast microwave switch is placed before the output of the system and is actuated throughout the pulse sequences. Similar switches are placed at the input of the box and provide 40~dB of attenuation when open to ensure that the high power pulses used for performing ESR do not reach the demodulator with a power exceeding its damage threshold. Power entering the box is further amplified before being homodyne demodulated. The resulting quadrature signals are digitized by a Data Acquisition System (DAQ). We route the pump through the box and use fast microwave switches to gate its output. Triggering of the AWG, DAQ, and microwave switches is achieved with TTL logic supplied by Spin-Core PulseBlaster ESR Pro.

\section{Device details}\label{sec:device}

\begin{figure}[ht]
	\centering
	\includegraphics[width=0.7\textwidth]{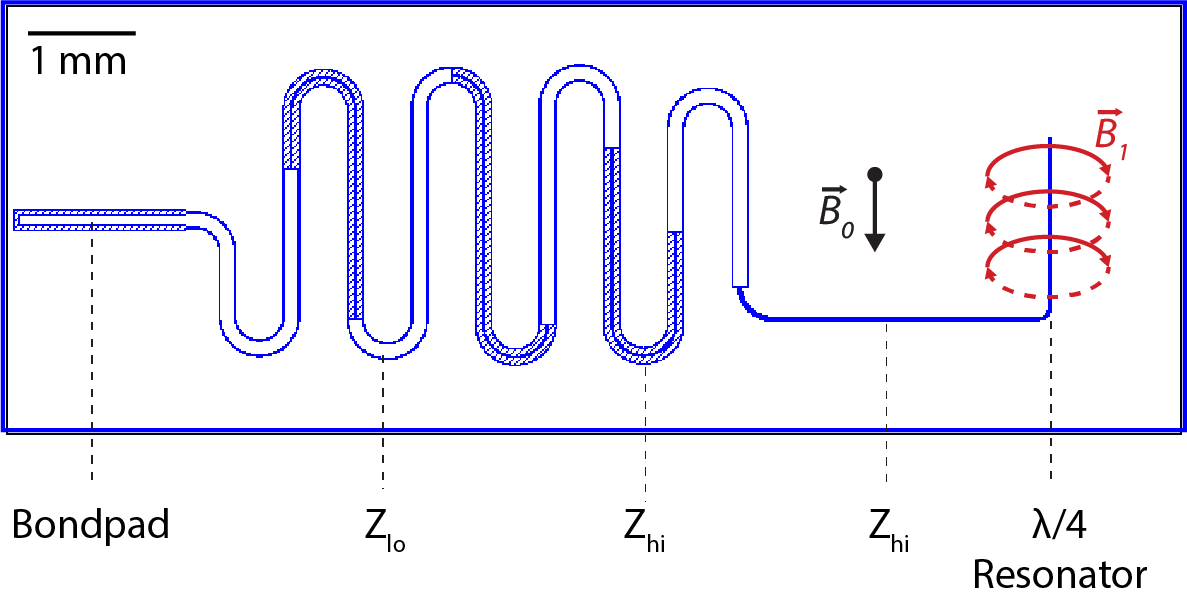}
	\caption[]{A top-down view of the device. The white areas correspond to a thin film of NbTiN. The blue areas have been etched away to reveal the Si underneath. All of the components are fabricated from a continuous CPW with the dimensions specified in Table~\ref{table-sm:device_design}. The orientation of the static $B_0$ field is in plane and parallel to the long axis of the resonator.}
	\label{fig-sm:device}
\end{figure}

A schematic of the device we use is shown in Fig.~\ref{fig-sm:device}. It consists of a $\lambda/4$ resonator formed from a CPW with centre track width $w=1~\mu$m and gap width $g=10~\mu$m, shorted to the ground plane on one end and galvanically connected to a Stepped Impedance Filter (SIF) on the other. It has a total of eight segments with alternating impedance ($4\times Z_\mathrm{lo}$ and $4\times Z_\mathrm{hi}$), each of which has an electrical length of $\lambda/4$ at $\omega_0$. The dimensions of the CPWs used for each section of the device are summarized in Table~\ref{table-sm:device_design}. The $B_0$ field is nominally aligned in plane and parallel to the long axis of the resonator. In this configuration, the resonant current produces a $B_1$ field that is perpendicular to $B_0$ for \Bi\ located underneath the centre conductor of the resonator.

\begin{table}[ht]
\begin{center}
\caption{Device design parameters.}
\begin{tabular}{c|c|c|c|c|c}
 & Bondpad & SIF $Z_\mathrm{lo}$ & SIF $Z_\mathrm{hi}$ & SIF $Z_\mathrm{hi}$ (final section) & Resonator \\
 \hline \hline
 $w$ ($\mu$m) & 100 & 138 & 10 &  5 & 1 \\
 $g$ ($\mu$m) & 45 & 6 & 70 & 15 & 10 \\
\end{tabular}
\label{table-sm:device_design}
\end{center}
\end{table}

To fabricate the device we begin with a $20~\mu$m thick epitaxial layer of \Sitwoeight\ (500~ppm residual {\Sitwonine}) grown on a $300~\mu$m high-resistivity non-compensation-doped natural-silicon wafer ($>5$~k$\Omega$). We implant \Bi\ donors uniformly across the wafer with a rectangular implantation profile with a target concentration of $10^{17}~$cm\textsuperscript{-3} from $0.35~\mu$m to $1.5~\mu$m depth (Fig.~\ref{fig-sm:implant_profile}). Damage to the Si lattice caused by the implantation is repaired via a 20~min anneal at 800~\degC\ in a {\Nitrogen} environment, which also aids in the incorporation of {\Bi} into the host crystal. It is expected that this procedure results in the electrical activation of approximately 60\% of the implanted donors \cite{weis2012electrical}. Following this, a $50~$nm film of NbTiN is sputtered (Star Cryoelectronics). From independent calibrations we determine that the kinetic inductance is $L_{k0}=3.5~\textrm{pH}/\square$. The resonator is patterned with EBL and etched with a CF\textsubscript{4} and Ar plasma.

The device is mounted within a $4\times 3~$mm\textsuperscript{2} groove in a Printed Circuit Board (PCB) with wax. The PCB is made from 0.635~mm thick Rogers RO3006 laminate covered with 1~oz of copper on both sides with an immersion silver finish. We use wirebonds to connect the PCB and device ground. We also place three wirebonds across the first six $\lambda/4$ segments of the SIF to prevent excitation of the slotline mode. The device is placed in a gold-plated oxygen-free copper enclosure. The body and lid of the enclosure are machined so that the 8~mm length of the resonator that is not supported by the PCB extends within a small 3D cavity. The cavity is designed to support a resonant mode at 30~GHz to suppresses radiation loss of the resonator.

\section{Estimating the sensitivity $N_\text{min}$}\label{sec:Nmin}
The sensitivity is defined as the minimum number of spins ($N_\text{min}$) required to produce a single spin echo with a $\snr = 1$. In order to determine $N_\text{min}$ for our spectrometer, we measure an echo signal to find the single-shot SNR and combine this with an estimate of the total number of spins ($N_\text{tot}$) that contribute to the echo signal. In the following section we provide detailed steps on how to calculate $N_\text{tot}$ for our device and provide an estimate for $N_\text{min}$ for the experiments we performed at $B_0=6.78~$mT.

\subsection{Implanted donors inside the magnetic mode volume}\label{sec:Nd}
\begin{figure}[ht]
	\centering
	\includegraphics[width=0.5\textwidth]{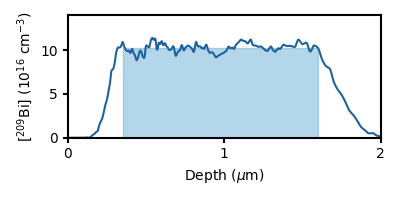}
	\caption[]{Donor implantation profile. The concentration of \Bi\ as a function of depth, simulated with the software package TRIM. The spins were implanted uniformly across the entire substrate. The shaded box corresponds to the rectangular profile used in our calculation of $N_\mathrm{min}$.}
	\label{fig-sm:implant_profile}
\end{figure}

Bismuth donors have been implanted implanted over the entire surface of the chip, with an implantation profile shown in Fig.~\ref{fig-sm:implant_profile}. However, only donors that interact with the magnetic mode of the resonator contribute to the echo signal. As a simple approximation, we could assume that the mode is confined to the $\lambda/4$ segment at the end of the device (designed to be the resonator), which consists of a coplanar waveguide with an inner track width of $w = 1~\mu$m and a length of $l = 1.75$~mm. The implantation profile is approximately rectangular, with a concentration of $C_d = 1\times 10^{17}~\text{cm}^{-3}$ from a depth of $0.35~\mu$m to $1.6~\mu$m. This provides an effective `donor volume' of $V_d \approx wl\Delta d = 2.2\times 10^{-15}~\text{m}^{3}$ (where $\Delta d = 1.25~\mu$m is the implantation depth range) and a number of implanted donors $N_d = \beta_aC_dV_d = 1.3\times 10^8$ able to interact with the resonator. This estimate accounts for the 60\% bismuth electrical activation yield ($\beta_a = 0.6$) under the annealing conditions used here \cite{weis2012electrical}.

This crude estimate of $N_d$ can be improved with detailed knowledge of the magnetic field distribution of the resonant mode in the device. This would allow us to calculate $N_d$ as

\begin{equation}\label{eq:Nd}
    \begin{aligned}
        N_d ={}& \frac{\int |B_{1\perp}(r)|^2\rho(r) dV}{\text{max}(|B_1|)^2},\\
        \approx{}& \beta_aC_d\frac{\int_\text{imp.} |B_{1\perp}(r)|^2 dV}{\text{max}(|B_1|)^2},\\
        ={}& \beta_aC_dV_d,
    \end{aligned}
\end{equation}

\noindent where $\rho(r)$ is the implanted and activated bismuth donor concentration, $B_{1\perp}(r)$ is the component of the $B_1$ field perpendicular to the static field $B_0$ and $\text{max}(|B_1|)$ is the maximum strength of the field over the whole mode. Using $B_{1\perp}(r)$ in the numerator of Eq.~\ref{eq:Nd} ensures that we only consider spins in the device which can be driven through the $\ket{F,m_f}=\ket{4,-4} \rightarrow \ket{5,-5}$ transition that we probe at $\sim 6.78$~mT. In the second line we assume that $\rho(r) = \beta_aC_d$ is constant over the depth range $0.35 - 1.6~\mu$m and zero otherwise, with the integration volume being restricted to this implantation depth range. We define the effective donor volume to be

\begin{gather}
    V_d = \frac{\int_\text{imp.} |B_{1\perp}(r)|^2 dV}{\text{max}(|B_1|)^2}
    \label{eq:Vd}
    \\
    = \underbrace{\left(\int \frac{|B_{1\perp}(r)|^2}{\text{max}(|B_1|)^2} dV \right)}_{V_m} \;\; \underbrace{\left(\frac{\int_\text{imp.}|B_{1\perp}(r)|^2 dV}{\int|B_{1\perp}(r)|^2 dV}\right)}_{\eta}
    \label{eq:Vd2}
,\end{gather}

\noindent which provides a more accurate measure than one based on simple geometric arguments presented above.  In the second line we show that $V_d$ can also be expressed as $V_d = V_m\eta$, where $V_m$ is the total volume of the perpendicular magnetic field component (the magnetic mode volume) and $\eta$ is the proportion of the perpendicular field occupied by the implanted spins (the spin filling factor).

\begin{figure}[ht]
	\centering
	\includegraphics[width=\textwidth]{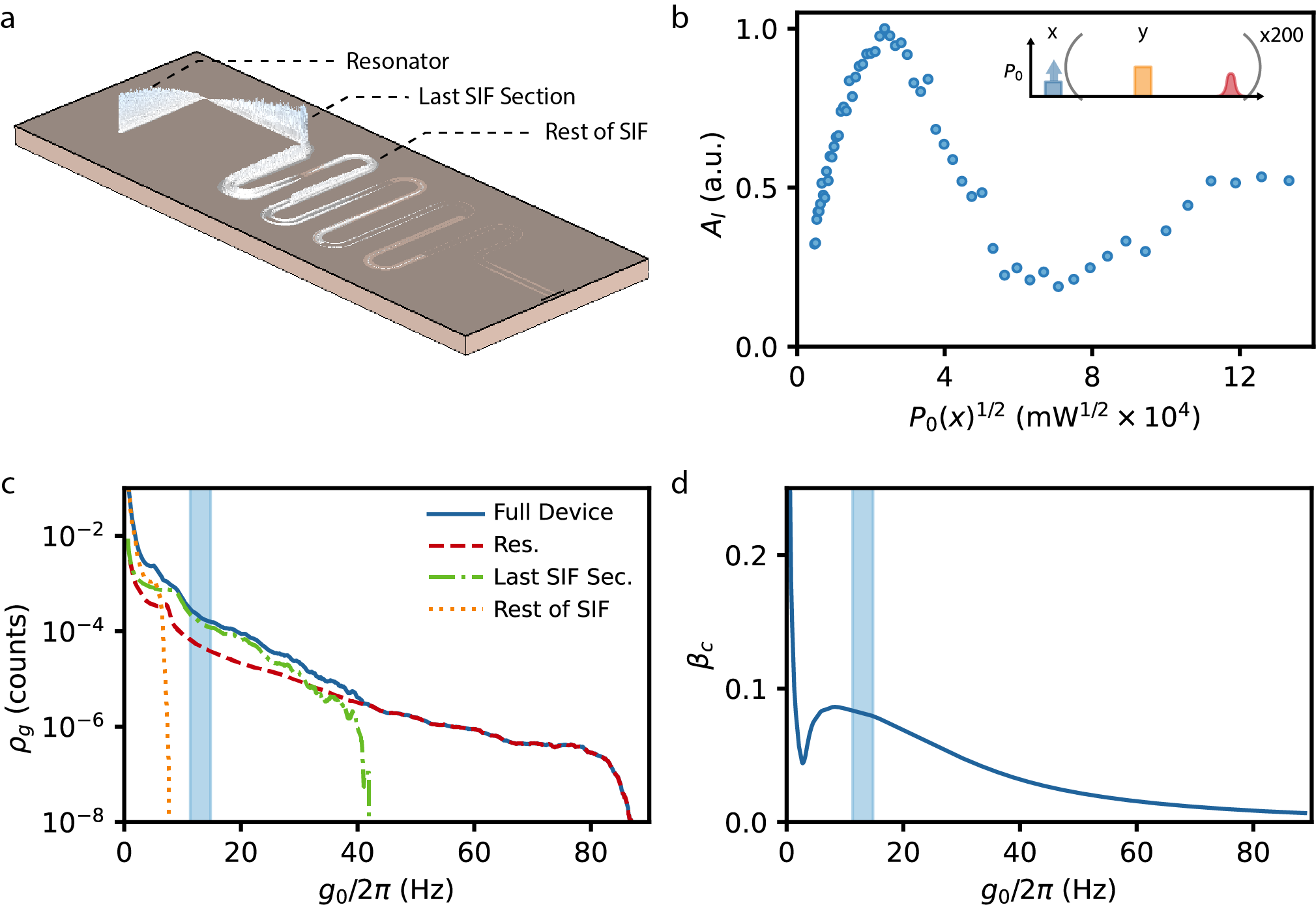}
	\caption[]{Distributions of spin-photon couplings. (a) A finite element simulation of the device. The strength of the magnetic field at $\omega_0$ is depicted as an out-of-plane gradient fill, and shows that the magnetic field is concentrated in the resonator and the last section of the SIF. (b) Rabi oscillations measured as a function of the tipping pulse power. Inset: a depiction of the CPMG-200 sequence used. (c) Normalized histograms depicting the number of total spins that couple to the magnetic mode of the device ($\propto \bar{V}_d$) as a function of $g_0$. (d) $\beta_c$, the ratio of total spins that contribute to a spin echo, as a function of $g_0$. In (c) and (d) the blue bars correspond to the range of $g_0$ extracted from the Rabi oscillations.}
	\label{fig-sm:g0_dist}
\end{figure}

To evaluate $B_1(r)$ we use CST Studio Suite to perform a finite-element simulation of the full planar device structure, incorporating kinetic inductance, bond wires and the device enclosure. We utilise the frequency domain solver to calculate the magnetic field profile of the fundamental resonator mode and re-scale the result to provide the root-mean-square (RMS) field fluctuations ($\delta B_{1\perp}(r)$). Calculating $\delta B_{1\perp}(r)$ has the dual purpose of allowing us to determine the single spin to resonator coupling strength ($g_0$), which we consider later in Section~\ref{sec:g0}. We find $\delta B_{1\perp}(r)$ using the relation

\begin{equation}\label{eq:brms}
    \delta B_{1\perp}(r) = \frac{B_{1\perp}(r)}{2\sqrt{\bar{n}}}
,\end{equation}

\noindent where $\bar{n}$ is the number of intracavity photons and is given by \cite{Bruno2015}

\begin{equation}\label{eq:nph}
    \bar{n} = \frac{4\kappa}{\hbar\omega_0(\kappa+\gamma)^2}P_\text{in}
.\end{equation}

\noindent Here $P_{in} = 0.5$~W is the input power used in the simulation, $\omega_0$ is the fundamental mode frequency, $\kappa=\omega_0/Q_L$ where $Q_L$ is the loaded quality factor, and $\gamma=\omega_0/Q_i$, where $Q_i$ is the internal quality factor. Fig.~\ref{fig-sm:g0_dist}a presents a depiction of the $\delta B_{1\perp}(r)$ distribution of the fundamental resonator mode.

Evaluating Eq.~\ref{eq:Vd} using our CST simulation result we find $V_d = 1~$pL, which using Eq.~\ref{eq:Nd} leads to $N_d = 6\times 10^{7}$. This value is smaller than the estimate derived from simple geometric arguments because the latter does not factor in the orientation of the $B_1$ field. This calculation is also more accurate because it takes into account the fact that the mode is not strictly confined to the last $\lambda/4$ segment, but instead penetrates into the stepped-impedance filter (see Fig.~\ref{fig-sm:g0_dist}a). Table~S\ref{table:vd} shows the effective donor volumes for the different regions in the device, where we find that approximately half of the mode exists in the last segment of the SIF.

\begin{longtable}[c]{| c | c |}
 \caption{Effective donor volumes throughout the device. $1~\mu \text{m}^3 = 1~$fL. \label{table:vd}}\\
 \hline
 Section & $V_d~(\mu\text{m}^3)$ \\
 \hline
 \endfirsthead

 \hline
 \endfoot
 \hline
 \endlastfoot
 SIF (first sections) & 189.5\\
 SIF (final section) & 549.8\\
 Resonator & 270.5\\
 \hline
 Total & 1009.8
\end{longtable}

\subsection{Coupling strength distribution}\label{sec:g0}
In microresonator devices the spin-photon coupling strength ($g_0$) is highly inhomogeneous due to the spatial distribution of the magnetic field. A spin echo measurement therefore only measures the subset of spins where the pulses provide well-calibrated $\pi$ and $\pi/2$ rotations \cite{ranjan2020pulsed,osullivan2020spin}. We must therefore also estimate what fraction of spins have coupling strengths that contribute to the echo signals. We do this by measuring $g_0$ via a Rabi oscillation experiment and simulating the coupling strength distribution in our device, as detailed below.

\subsubsection{Experiment}
We can extract an average $g_0$ for the spins experimentally by measuring the Rabi frequency ($\Omega_R$) along with the average number of intracavity photons ($\bar{n}$) generated by the input pulse. They are related by \cite{Haroche2006}:

\begin{equation}\label{eq:Rabi}
    \Omega_R = 2g_0\sqrt{\bar{n}}
,\end{equation}

\noindent where $\bar{n}$ is calculated from Eqn.~\ref{eq:nph}.

In Fig.~S\ref{fig-sm:g0_dist}b we perform Rabi oscillations by fixing the durations of the $\pi$ and $\pi/2$ pulses ($t_\mathrm{pulse}=3~\mu$s) in a CPMG-200 measurement sequence and varying the power of the $\pi/2$ pulse. We assign the power at which $A_I$ is maximum to be the $P_0(\pi/2)$, which in this case was $-73\pm 1~$dBm, referred to the input of the device. This power is determined by subtracting the loss between the room temperature microwave source and the sample, which was calibrated by combining measurements made across a series of cool downs of our cyrostat. We estimate $\bar{n} = 1.00_{0.80}^{1.26}\times 10^7$ from VNA measurements of $S_{11}$ measured at the same power as $P_0(\pi/2)$, where the upper and lower values correspond to the uncertainty in $P_0$. We therefore extract $g_0/2\pi = 13.2_{11.7}^{14.8}~$Hz.

\subsubsection{Simulation}
Our CST Suite simulation of $\delta B_{1\perp}$ (see Sec.~\ref{sec:Nd}) can be used to calculate the distribution of spin-resonator coupling strengths using the formula:

\begin{equation}\label{eq:g0}
    g_0 = \delta B_{1\perp}M\gamma_e
,\end{equation}

\noindent where $M=0.473$ is the matrix element for the $\ket{F,m_f}=\ket{4,-4} \rightarrow \ket{5,-5}$ transition at $B_0=6.78~$mT and $\gamma_e/2\pi = 28$~GHz/T is the gyromagnetic ratio. In Fig.~S\ref{fig-sm:g0_dist}c we show a normalized histogram of the number of spins that couple to the resonant mode ($\rho_g$) as a function of $g_0$. Comparing the histograms for different parts of the device it becomes apparent that the SIF, which contains larger CPW dimensions than the rest of the structure, has the weakest coupling strengths. The last segment of the SIF and the resonator produce much larger couplings, due to the tighter confinement of the magnetic field in these regions. The range of $g_0$ corresponding to our experimental measurement are indicated by a blue bar, where the ratio of spins located underneath the last section of the SIF and the resonator is approximately $3:1$.

\subsubsection{Spin fraction}
Using the simulated distribution of $g_0$ in our device and the experimentally extracted $g_0$, we can estimate the fraction of resonant spins that contribute to an echo. We define the spin fraction as \cite{ranjan2020pulsed}:

\begin{equation}\label{eq:betac}
    \beta_c(g_0) = \frac{\int{g\rho_g(g)\sin^3{\left(\frac{\pi g}{2g_0}\right)} dg}}{\int{g\rho_g(g) dg}}
,\end{equation}

\noindent and plot $\beta_c(g_0)$ in Fig.~S\ref{fig-sm:g0_dist}d. Over the range of $g_0$ extracted from our measurement (indicated by a blue bar) we find $\beta_c=0.082_{0.079}^{0.083}$.

\subsection{Pulse excitation bandwidth}
Because we use a high-$Q$ resonator, we must also consider that the bandwidth of the spins we excite in pulsed measurements is smaller than the bandwidth of the ESR transitions. It is therefore necessary to find the overlap of the pulse excitation bandwidth with the spectral density of a given ESR transition to estimate the proportion of spins that are excited.

 \begin{figure}[ht]
 \centering
 \includegraphics[width=0.7\textwidth,keepaspectratio]{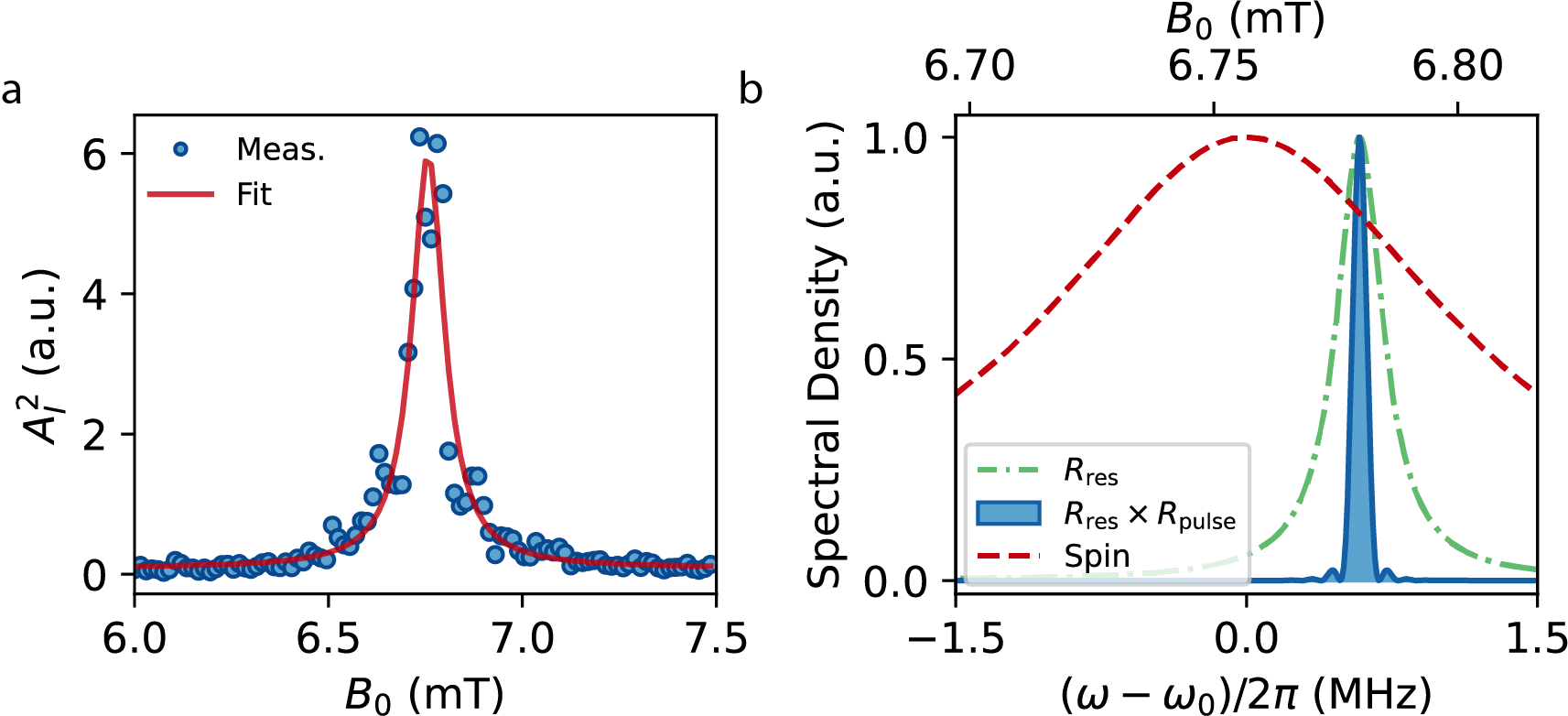}
 \caption[Pulse Excitation Bandwidth]{Spin and resonator spectral density. (a) A measurement of the $\ket{F,m_f}=\ket{4,-4} \rightarrow \ket{5,-5}$ transition using a CPMG sequence with 200 $\pi$ pulses. The data is fit with a Lorentzian from which we extract a linewidth of 0.10~mT. (b) The magnitude response of the cavity (dashed green line) and the cavity-filtered pulse power spectrum (blue fill) are considerably narrower than the spin linewidth (dashed red line).}
 \label{fig-sm:pulse_excitation_bandwidth}
 \end{figure}

In Fig.~\ref{fig-sm:pulse_excitation_bandwidth}a we measure a \Bi\ ESR transition centred at $B_0=6.76~$mT using a Carr-Purcell-Meiboom-Gill (CPMG) sequence with 200 refocusing $\pi$ pulses as we sweep $B_0$. Previous investigations of \Bi\ ESR transitions in Si using high-$Q$ aluminium resonators have found broad spin distributions that were non-uniform with $B_0$ \cite{bienfait2016reaching}. For those devices, the transitions were determined to be broadened by local strain, caused by the larger thermal contraction of the aluminium resonator compared to the silicon substrate \cite{pla2018strain, ranjan2021spatially}. Since the ESR transitions we observe in our device are similarly broad, this strongly suggests a similar effect of local strain in our device. Here we estimate the ratio of spins that lie within the bandwidth of our measurement by comparing the spectral density of the spin transition with the power spectrum of the pulses employed in the measurements.

We define the spin spectral density as $A_I^2(B_0) = \int_{T_E}{I^2 dt}$, where $T_E$ defines the spin echo duration. From numerical calculations of the \Bi\ spin Hamiltonian we determine that the ESR transition resonant with our device is the $\ket{F,m_f}=\ket{4,-4} \rightarrow \ket{5,-5}$ transition, which has $(\partial \omega/\partial B)/2\pi= -25.06~$MHz/mT. This allows us to convert $A_I^2$ from a function of $B_0$ to a function of $\omega-\omega_0$. We note that for a spin ensemble inhomogenously broadened by strain, there is no strict functional form expected for the lineshape \cite{pla2018strain, ranjan2021spatially}, but for this experiment a Lorentzian seems to fit the data well. This therefore allows us to estimate the spin linewidth to be $0.10~\text{mT}$, or equivalently $2.55$~MHz.

The bandwidth of spins that are excited by a pulse ($R$) is given by the product of the magnitude response of the resonator ($R_\text{res}$) and the pulse power spectrum ($R_\text{pulse}$)

\begin{equation}
    \begin{aligned}
        R_\text{res}(\omega) ={}& \frac{1}{1 + 4(\omega-\omega_0)^2/\kappa^2},\\
        R_\text{pulse}(\omega) ={}& 4 \text{sinc}^2\left[\frac{t_p(\omega - \omega_0)}{2}\right],\\
        R(\omega) ={}& R_\text{res}R_\text{pulse},
    \end{aligned}
\end{equation}

\noindent where $\kappa=\omega_0/Q_L$, $\omega_0/2\pi = 7.203$~GHz, $Q_L$ is the loaded quality factor of the resonator mode and $t_p$ is the fixed duration of the $\pi/2$ and $\pi$ pulses used in our measurements. We use $Q_i= 117 \times 10^3$ and $Q_L=(Q_i^{-1} + Q_c^{-1})^{-1} = 28.2\times 10^3$, which are extracted from a fit of VNA measurements of $S_{11}$ at the same power as the pulses used in our measurements. For our experiments $t_p=10~\mu$s, which sets the limit to the excitation bandwidth.

In Fig.~S\ref{fig-sm:pulse_excitation_bandwidth}b we compare $A_I^2(\omega)$, $R_\text{res}(\omega)$ and $R(\omega)$, which are all normalized. We note that $R_\text{res}(\omega)$ and $R(\omega)$ are offset from zero detuning (which we reference to $A_I^2$) because the measurements were completed at $6.78$~mT, whereas the fit gives $B_0=6.76~$mT as the center of the transition. The ratio of spins within the $\ket{F,m_f}=\ket{4,-4} \rightarrow \ket{5,-5}$ transition that we excite in our pulsed measurements is then estimated as

\begin{equation}
\beta_b = \frac{\int R(\omega)A_I^2(\omega)d\omega}{\int A_e^2(\omega)d\omega} \approx \frac{1}{54.3}
.\end{equation}

\subsection{Total number of spins}
We are now able to estimate that the total number of spins coupled to the mode of the microwave resonator that contribute to the detected spin echo signals is $N_\text{tot} = \beta_b \beta_c \beta_d N_d$. Here, $\beta_d = 1/10$ accounts for the ten equally populated ground states of the bismuth donor spin system at the measurement temperature ($400~\text{mK}$) and magnetic field ($B_0 \sim 6.78~\text{mT}$). Therefore:

\begin{equation}
N_\mathrm{tot} = \beta_a\beta_b\beta_c\beta_d C_d V_d \approx 9.4 \times 10^3
.\end{equation}

\noindent The definition and numerical values for each of the quantities defining $N_\mathrm{tot}$ are summarized in Table~\ref{table-sm:Ntot_defn}. 

\begin{longtable}[c]{| c | c | l |}
 \caption{Summary of factors defining $N_\mathrm{tot}$. \label{table-sm:Ntot_defn}}\\

 \hline
 & Value & Description \\
 \hline
 \endfirsthead

 \hline
 \endfoot
 \hline
 \endlastfoot
    $\beta_a$ & 0.6 & Donor activation ratio \\
    $\beta_b$ & 1/54.3 & Ratio of cavity-pulse bandwidth to total spin linewidth \\
    $\beta_c$ & 0.082 & Fraction of resonant spins with the appropriate coupling $g_0$ \\
    $\beta_d$ & 1/10 & Fraction of spins at the measured spin transition \\
    $C_d$ & $1.03 \times 10^{17}~\text{cm}^{-3}$ & Donor concentration \\
    $V_d$ & $1009.8~\mu\text{m}^3$ & Effective donor volume \\
\end{longtable}

\subsection{Sensitivity estimate}
\label{section:sensitivity_estimate}

The sensitivity is defined as the minimum number of spins needed to produce a single spin echo with a SNR of unity. To estimate $N_\text{min}$, we determine the single-shot amplitude SNR for a Hahn echo ($\snr_1$) with different applied amplifier gains. We then define $N_\text{min} = N_\text{tot}/\snr_1$. The results are displayed in Table~\ref{table:Nmin}.

\begin{longtable}[c]{| c | c | c | c | c |}
 \caption{Single-shot echo SNR and spin sensitivity versus device amplitude gain. Experiments one and two were performed on different days. \label{table:Nmin}}\\
 \hline
 & \multicolumn{2}{c}{Experiment 1} & \multicolumn{2}{c}{Experiment 2} \\
 \hline
 KIPA Amplitude Gain (dB) & $\snr_1$ & $N_\text{min} \times 10^{-3}$ & $\snr_1$ & $N_\text{min} \times 10^{-3}$ \\
 \hline
 \endfirsthead

 \hline
 \endfoot
 \hline
 0 & 0.55(10) & 17.1 & 0.42(0.12) & 22.4 \\
 2.1 & 1.91(13) & 4.9 & 1.50(0.16) & 6.3 \\
 4.4 & 2.83(32) & 3.3 & - & - \\
 5.6 & - & - & 2.61(0.39) & 3.6 \\
 8.0 & 3.99(33) & 2.4 & 3.36(0.41) & 2.8 \\
\end{longtable}

Therefore at the highest gain, we are sensitive to approximately to 2,800 spins in a single shot measurement. A theoretical estimate of the sensitivity is provided by \cite{bienfait2016reaching}

\begin{equation}\label{eq:Nmin}
    N_\text{min} = \frac{\kappa+w}{2g_0p}\sqrt{\frac{n_nw\kappa}{\kappa_c(\kappa+2w)}}.
\end{equation}

\noindent where $\kappa_c = \omega_0/Q_c$ is the coupling rate to the external port, $w^{-1} = T_E$ is the spin echo duration, $p = \tanh{[\hbar\omega_0/(2k_BT)]}$ is the spin polarization (where $k_B$ is the Boltzmann constant and $T$ the temperature) and $n_n$ is the number of noise photons added by the detection chain. This equation has been verified in a number of experiments and shown to produce reliable estimates \cite{bienfait2016reaching,ranjan2020electron,ranjan2020pulsed}. Using our calculations of $N_\text{min}$ and measured values for the resonator bandwidth, spin polarization $p = 0.40$, $T_E \approx 20~\mu$s and $g_0/2\pi = 13.2$~Hz, we can also estimate $n_n$. With the pump off (no device gain), the first amplifier in the chain is a high-electron-mobility-transistor (HEMT) amplifier, and we calculate $n_n \approx 14$ photons. Our commercial HEMT amplifier (LNF-LNC0.3\_14B) has a specified noise temperature of $T_n = 3.6$~K. Combining this with the estimated 3.5~dB of insertion loss between the HEMT and device (measured at room temperature), we can provide a rough estimate the system noise that is added to the spin echo signals as $n_n = (1/4 + n_\text{th})/\eta \approx 12$~photons, where $n_\text{th} = 1/(2[\exp(\hbar\omega_0/k_BT_n)-1])$ is the thermal noise per signal quadrature and $\eta = 10^{(-3.5/10)} = 0.45$. At the highest amplitude gain of 8~dB, our extracted $N_\text{min} = 2,800$ provides an estimated noise of $n_n = 0.22$~photons, whereas the equilibrium noise is expected to be $n_n = 1/4 + n_\text{th} = 0.6$~photons at $T = 400$~mK, in close agreement.

\section{Scaling of the SNR with amplifier gain}
\label{sec:SNR_vs_Gk}

The scaling of the SNR with the gain of the KIPA can be explained using a model based on cavity input-output theory. Here we describe the model and explain how it is used to fit our experimental data.

\subsubsection{Input-output model}
\begin{figure}[!ht]
\includegraphics[width=0.4\textwidth]{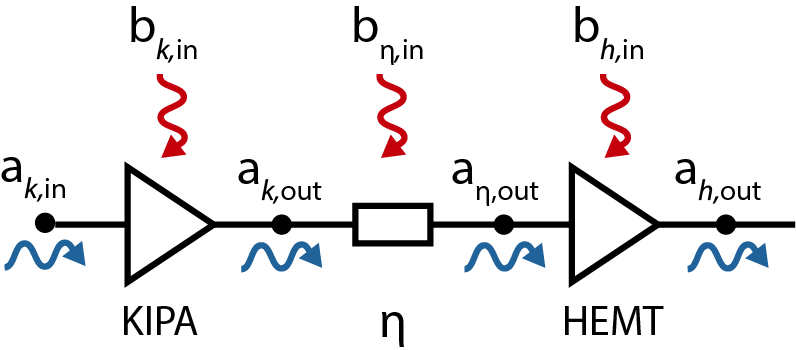}
\caption[]{A model of the measurement system based on cavity input-output theory. The blue arrows label the signal mode as it passes through the three components. The red arrows correspond to bath modes that add noise to the signal.}
\label{fig-sm:input_output_model}
\end{figure}

We consider that the KIPA is connected in series with an attenuator with power transmissivity $\eta^2$ and a HEMT (Fig.~\ref{fig-sm:input_output_model}). The mode propagating into each component is given by $a_{j,\mathrm{in}}$, where $j\in (k,\eta,h)$ is a label for each of the circuit components. Each mode $a_{j,\mathrm{in}}$ has the corresponding quadrature operators

\begin{gather}
I_{j,\mathrm{in}} = \frac{a_{j,\mathrm{in}} + a^\dag_{j,\mathrm{in}}}{2},
\label{eqn_Xj_in} \\
Q_{j,\mathrm{in}} = \frac{a_{j,\mathrm{in}} - a^\dag_{j,\mathrm{in}}}{2i}
\label{eqn_Yj_in}
.\end{gather}

\noindent Each component also has output modes $a_{j,\mathrm{out}}$, with quadrature operators equivalent to Eqs.~\ref{eqn_Xj_in} and \ref{eqn_Yj_in}, where the subscripts `in' are replaced with `out.' Because the circuit is linear, $a_{k,\mathrm{out}}=a_{\eta,\mathrm{in}}$, i.e. the mode entering the attenuator is simply the mode that exits the KIPA, and so forth. We also consider bath modes connected to each circuit component, $b_{j,\mathrm{in}}$, that add noise to the signal. The bath modes have the associated quadrature operators

\begin{gather}
I_{bj,\mathrm{in}} = \frac{b_{j,\mathrm{in}} + b^\dag_{j,\mathrm{in}}}{2},
\\
Q_{bj,\mathrm{in}} = \frac{b_{j,\mathrm{in}} - b^\dag_{j,\mathrm{in}}}{2i}
.\end{gather}

When amplifying in degenerate mode (i.e. $\omega_p = 2\omega_0$), the input-output relations for the KIPA are given by \cite{Parker2021}

\begin{gather}
\begin{pmatrix} I_{k,\mathrm{out}} \\ Q_{k,\mathrm{out}} \end{pmatrix}  = 
A_G \begin{pmatrix} I_{k,\mathrm{in}} \\ Q_{k,\mathrm{in}} \end{pmatrix} + \sqrt{\frac{\gamma}{\kappa}} (A_G + 1)\begin{pmatrix} I_{bk,\mathrm{in}} \\ Q_{bk,\mathrm{in}} \end{pmatrix}
,\end{gather}

\noindent where

\begin{equation}
A_G = \frac{\kappa}{\Delta^2 + (\kappa + \gamma)^2/4 - |\zeta|^2}\begin{pmatrix}
(\kappa + \gamma)/2 - |\zeta|\sin(\phi_p) & -|\zeta|\cos(\phi_p) + \Delta \\ -|\zeta|\cos(\phi_p) - \Delta & (\kappa + \gamma)/2 + |\zeta|\sin(\phi_p)
\end{pmatrix} - 1
.\end{equation}

\noindent Here, $\kappa$ is the rate at which the intracavity field is coupled to the measurement port, $\gamma$ is the rate at which the intracavity field is coupled to internal losses, $\Delta=(\omega_0 - \omega_p/2)$ is the detuning of the pump from twice the cavity's resonant frequency, and $\zeta$ is the three-wave mixing strength. For $\Delta=\cos(\phi_p)=0$, $A_G$ is diagonal. Given that a spin echo signal can be aligned along a single quadrature, we therefore focus on how $I_{k,\mathrm{in}}$ is transformed to $I_{k,\mathrm{out}}$ when $\Delta=0$ and $\phi_p=3\pi/2$, which maximizes the gain along the $I$-quadrature. We find

\begin{gather}
    I_{k,\mathrm{out}} = G_k I_{k,\mathrm{in}} + \sqrt{\frac{\gamma}{\kappa}}(G_k + 1)I_{bk,\mathrm{in}}
    \label{eqn-sm:input_output_kipa_I}
\end{gather}

\noindent where

\begin{equation}
    G_k = \frac{\kappa}{(\kappa+\gamma)/2 - |\zeta|} - 1
.\end{equation}

To find how the signal is transformed, we take the expectation values of both sides of Eq.~\ref{eqn-sm:input_output_kipa_I} where we assume that the bath mode is a thermal state (i.e. $\braket{I_{bk,\mathrm{in}}}=0$), which yields

\begin{equation}
    \braket{I_{k,\mathrm{out}}} = G_k \braket{I_{k,\mathrm{in}}}
.\end{equation}

To find how the noise is transformed, we calculate the expectation value $\braket{I^2_{k,\mathrm{out}}}$, which is equal to \cite{Parker2021}

\begin{equation}
    \braket{I^2_{k,\mathrm{out}}} = G_k^2 \braket{I^2_{k,\mathrm{in}}} + (G^2_k - 1)n_k
,\end{equation}

\noindent where 

\begin{equation}
    n_k = \frac{\gamma}{\kappa} \frac{G_k + 1}{G_k - 1} \braket{I^2_{bk,\mathrm{in}}} = \frac{\gamma}{\kappa} \frac{G_k + 1}{G_k - 1} \left(\frac{1}{4} + n_{k,\mathrm{th}}\right)
.\end{equation}

\noindent $n_k$ is the noise added by the KIPA to the $I$-quadrature (referred to its input) which has two components, vacuum noise and thermal noise ($n_{k,\mathrm{th}}$).

The input-output relation for the attenuator with amplitude transmissivity $\eta$ is given by

\begin{gather}
a_{\eta,\mathrm{out}} =  \eta a_{\eta,\mathrm{in}} + \sqrt{1-\eta^2} b_{\eta,\mathrm{in}}
\label{eqn-sm:input_output_eta}
,\end{gather}

\noindent where

\begin{gather}
\braket{I_{b\eta,\mathrm{in}}}=0,
\\
\braket{I^2_{b\eta,\mathrm{in}}} = \frac{1}{4} + n_\eta
.\end{gather}

\noindent Here $n_\eta$ is the noise power for each quadrature of the bath mode measured in excess of vacuum. We therefore see that the attenuator acts as a beamsplitter which mixes in noise with power $\propto 1-\eta^2$.

The HEMT is a non-degenerate amplifier and therefore has input-output relations specified by Caves' Theorem 

\begin{equation}
a_{h,\mathrm{out}} = G_h a_{h,\mathrm{in}} + \sqrt{G_h^2 - 1} b_{h,\mathrm{in}}^\dag
\label{eqn-sm:input_output_hemt}
,\end{equation}

\noindent where

\begin{gather}
\braket{I_{bh,\mathrm{in}}}=0,
\\
\braket{I^2_{bh,\mathrm{in}}} = \frac{1}{4} + n_h
.\end{gather}

\noindent $n_h$ is therefore the noise added by the HEMT in excess of vacuum to each of the quadratures. 

Combining Eqns.~\ref{eqn-sm:input_output_kipa_I}, \ref{eqn-sm:input_output_eta} and \ref{eqn-sm:input_output_hemt} we find

\begin{gather}
I_{h,\mathrm{out}} = G_hG_k\eta I_{k,\mathrm{in}} + G_h\sqrt{G_k^2-1}\eta I_{bk,\mathrm{in}} + G_h\sqrt{1-\eta^2} I_{b\eta,\mathrm{in}} + \sqrt{G_h^2-1}I_{bh,\mathrm{in}}
.\end{gather}

This leads to the expectation values

\begin{gather}
\braket{I_{h,\mathrm{out}}}^2 = G_h^2G_k^2\eta^2\braket{I_{k,\mathrm{in}}}^2,
\\
\braket{I^2_{h,\mathrm{out}}} = G_h^2G_k^2\eta^2\braket{I^2_{k,\mathrm{in}}} + G_h^2(G_k^2-1)\eta^2 n_k + G_h^2\eta^2 n_\mathrm{sys}
,\end{gather}

\noindent where $n_\mathrm{sys} = (1/\eta^2-1)(1/4 + n_\eta) +(1-1/G_h^2)(1/4 + n_h)/\eta^2$ is the `system noise' referred to the output of the KIPA. These expressions give the square of the signal and the square of the noise measured at the output of the HEMT. They can be referred to the input of the HEMT by dividing by $G_h^2$. Doing so and taking their ratio, we arrive at an expression for the SNR

\begin{gather}
(\mathrm{SNR})^2 = \frac{G_k^2 \braket{I_{k,\mathrm{in}}}^2}{G_k^2\braket{I^2_{k,\mathrm{in}}} + (G_k^2-1)n_k  + n_\mathrm{sys}}
\label{eqn-sm:SNR_scaling}
.\end{gather}

Eq.~\ref{eqn-sm:SNR_scaling} tells us that the SNR is limited by three sources of noise: noise on the signal at the input of the KIPA, noise added by the KIPA, and noise added by the components following the KIPA. We should therefore expect the SNR to grow with $G_k^2$ only so long as the first two contributions, which are amplified by the KIPA, remain smaller than or of a similar magnitude to $n_\mathrm{sys}$. Thereafter, both the signal and noise would amplified by an equivalent amount. We can see this clearly by calculating $G_\mathrm{SNR}^2$, which is given by

\begin{gather}
G_\mathrm{SNR}^2 = \frac{(\mathrm{SNR})^2}{\left.(\mathrm{SNR})^2\right|_{G_k^2=1}} = \frac{G_k^2(\braket{I^2_{k,\mathrm{in}}} + n_\mathrm{sys})}{G_k^2\braket{I^2_{k,\mathrm{in}}} + (G_k^2-1)n_k  + n_\mathrm{sys}}
\label{eqn_io_GSNR_scaling}
.\end{gather}

\noindent In the limit $G_k^2 \gg 1$, this simplifies to

\begin{gather}
G_\mathrm{SNR}^2 \approx \frac{\braket{I^2_{k,\mathrm{in}}} + n_\mathrm{sys}}{\braket{I^2_{k,\mathrm{in}}} + n_k}
\label{eqn_io_GSNR_high_gain_limt}
,\end{gather}

\noindent where we see that $G_\mathrm{SNR}$ is constant. Provided $n_\mathrm{sys} \gg \braket{I^2_{k,\mathrm{in}}}$, then $G^2_\mathrm{SNR}$ is approximately equal to the ratio of the noise added by the components following the KIPA and the noise at the KIPA's input. Therefore, the SNR will be improved by amplifying with the KIPA whenever $n_k < n_\mathrm{sys}$.

\subsubsection{Comparing the model to experiment}

In Fig.~4a of the main text we report the SNR of two sets of measurements made with $\phi_p=0$ as a function of the degenerate amplifier amplitude gain, $G_\mathrm{k}$. To independently match the input-output model to these experiments would require knowledge of all the circuit parameters, including: $G_h$, $\eta$, $n_\mathrm{sys}$, $n_k$ and $\braket{I^2_{k,\mathrm{in}}}$. This can only be achieved via a full analysis of the system noise temperature using a calibrated noise source, which is beyond the scope of the present study. Nevertheless, we can fit the experimental data to a generalized version of Eqn.~\ref{eqn-sm:SNR_scaling} to confirm that the model is consistent with our measurements. The data in Fig.~4a was fit with the function

\begin{gather}
\mathrm{SNR} = \sqrt{\frac{G_k^2 A}{G_k^2 B + 1}}
\label{eqn_io_GSNR_generalized}
,\end{gather}

\noindent where $A = \braket{I_{k,\mathrm{in}}}^2/(n_\mathrm{sys} - n_k)$ and $B = (\braket{I^2_{k,\mathrm{in}}} + n_k)/(n_\mathrm{sys} - n_k)$. The fitting parameter $B$ provides an estimate of the ratio of the noise at the input of the KIPA and the system noise, provided $n_\mathrm{sys} \gg n_k$. From the fit we extract $1/B=22.1$. As explained in Section~\ref{section:sensitivity_estimate}, we expect the input noise to the KIPA to be of order one photon (for two quadratures), and the system noise to be of order twenty photons (two quadratures). This result is therefore consistent with our expectations and the sensitivity estimate. We reiterate that this analysis is not intended to serve as a measurement of the noise temperature of the KIPA or the HEMT, but it does provide a reasonable sanity check. 

\section{Spin relaxation time}\label{sec:T1}

\begin{figure*}[!ht]
\includegraphics[width=\textwidth]{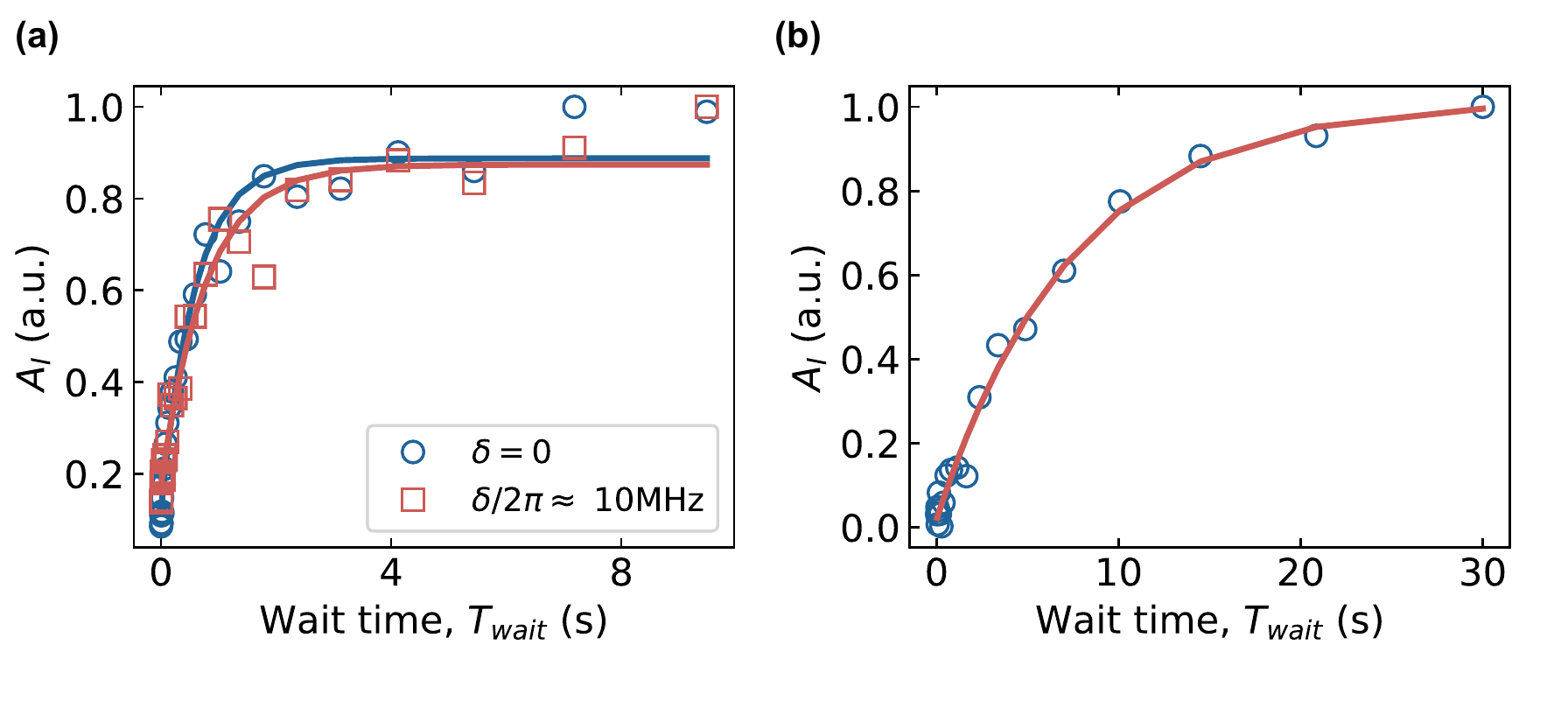}\label{fig:T1}
\caption[$T_1$ measurements]{Measurements of the longitudinal relaxation time $T_1$ for the electron spins. Integrated echo quadrature signal $A_{I}$ is plotted as a function of the wait time $T_\text{wait}$ after the spin magnetization is initially disturbed by the first pulse of a sequence. The signal is measured with a $N = 200$ pulse CPMG sequence. (a) Inversion recovery experiment performed when spin-cavity detuning $\delta$ is zero and when $\delta/2\pi\approx10~$MHz. Fits with exponential functions reveals $T_\text{1, on} = 600~$ms and $T_\text{1, off}=780~$ms when $\delta=0$ and $\delta\approx10~$MHz, respectively. (b) Saturation recovery experiment performed when $\delta =0$. The exponential fit yields $T_1=7.5~$s.}
\end{figure*}

Here we study the longitudinal relaxation processes, which are described by the characteristic time $T_1$. We again focus on the ESR transition $|F, m_F\rangle = |4, -4\rangle \rightarrow |5, -5\rangle$, which we measure at $6.78~$mT.
  
A known spin relaxation process in micro-resonator devices occurs due to the Purcell-enhanced spontaneous emission of microwave photons \cite{Bienfait2016d, ranjan2020pulsed}. This process is suppressed quadratically with the frequency detuning $\delta = \omega_0 - \omega_s$ between the cavity and the spin system according to $\Gamma_{p} = (\kappa + \gamma) g_{0}^2/[(\kappa + \gamma)^2/4 + \delta^2]$. To test if our electron spin energy relaxation rate is Purcell limited, we performed an inversion recovery experiment to measure $T_1$ when spins are on-resonance with the cavity ($\delta = 0$) and when they are detuned ($\delta/2\pi \approx10~$MHz) during the hold time $T_\text{wait}$. The spin-resonator detuning is realized by reducing the DC bias current in the resonator from $I_\text{DC} = 3.3~$mA to $I_\text{DC} = 2.8~$mA. The spin magnetization is measured after the wait period using a $N = 200$ CPMG pulse sequence with an inter pulse duration of $\tau = 75~\mu$s. All pulses had a duration of $t_p = 10~\mu$s. The sequence was repeated 3 times with a repetition of $T_\text{rep} = 30~$s. The integrated echo quadrature $I$ signals are presented in Fig.~S\ref{fig:T1}a as a function of $T_\text{wait}$. The echo decays are well fitted to an exponential function, allowing us to extract the longitudinal relaxation time $T_\text{1, on}\approx600(60)~$ms when spins are resonant with the cavity and $T_\text{1, off}\approx780(90)~$ms when cavity is detuned from the spins, confirming that the spin energy relaxation rate is not Purcell limited.

To reveal the underlying relaxation mechanism, we perform a saturation recovery experiment with the spins on-resonance with the cavity $\delta = 0$. The sequence commences with a $5~$s long saturation pulse, followed by the wait period $T_\text{wait}$ and spin magnetization measurement using a CPMG pulse sequence. The integrated echo quadrature response $A_{I}$ and associated exponential fit is presented in Fig.~\ref{fig:T1}b, indicating a $T_{1}\approx7.5~$s. The saturation recovery sequence \cite{schweiger2001principles} is known to suppress polarization mixing mechanisms, such as spectral and spatial spin diffusion, by reducing spin polarization gradients in the sample. Such mechanisms are also common in microresonator devices involving donors in silicon \cite{Bienfait2016d}. Comparing the values of $T_1$ obtained in the inversion recovery and saturation recovery measurements, it is clear that the dominant relaxation mechanism for our spin system is spin diffusion.

\section{Optimizing the signal-to-noise ratio}
\label{sec:optimizing_SNR}

Kinetic inductance is the crucial ingredient in our device that allows the in situ parametric amplification of spin echo signals. However, this inductance does not couple to spins and thus reduces the initial echo signal collected from the donor ensemble. Understanding how the kinetic inductance fraction $\alpha = L_k/(L_k+L_g)$ (where $L_k$ is the kinetic inductance and $L_g$ is the geometric inductance) affects the signal-to-noise ratio and other important parameters, such as the pump power, is therefore important and allows us to identify optimized device parameters.

\subsection{Pump power}
We wish to understand the dependence of the pump power needed to produce a specific gain on $\alpha$. The Hamiltonian for a kinetic inductance parametric amplifier, as utilized in this work, was derived in Ref.~\citenum{parker2021near}. In a frame rotating at half of the pump frequency $\omega_p/2$, the Hamiltonian is:
\begin{equation}\label{eq:HKIPA}
    \begin{aligned}
	    H_\kir ={}& \hbar\left(\omega_0 + \delta_\dc + \delta_p + K -\frac{\omega_p}{2}\right) a\dg a + \frac{\hbar\xi}{2}a{\dg}^2 + \frac{\hbar\xi^*}{2}a^2 + \frac{\hbar K}{2}a{\dg}^2 a^2,
    \end{aligned}
\end{equation}
with the following important Hamiltonian parameters defined as:
\begin{subequations}\label{eq:Hparam}
    \begin{align}
        \delta_\dc ={}& -\frac{1}{2}\frac{I_\dc^2}{I_*^2}\omega_0,\\
        \delta_p ={}& -\frac{1}{8}\frac{I_p^2}{I_*^2}\omega_0,\\
        K ={}& - \frac{3}{8}\frac{\hbar\omega_0}{L_TI_*^2}\omega_0,\\
        \xi ={}& -\frac{1}{4}\frac{I_\dc I_p}{I_*^2}\omega_0e^{-i\varphi_p}.
    \end{align}
\end{subequations}

This Hamiltonian was derived in the high kinetic inductance limit, where $\alpha \approx 1$. We have re-derived Eq.~\ref{eq:Hparam} assuming a non-negligible geometric inductance (i.e. $\alpha < 1$) and find the Hamiltonian to be identical, but with the parameters scaled by the factor $\alpha$:
\begin{subequations}\label{eq:Hparam2}
    \begin{align}
        \delta_\dc ={}& -\frac{\alpha}{2}\frac{I_\dc^2}{(I_*[\alpha])^2}\omega_0,\\
        \delta_p ={}& -\frac{\alpha}{8}\frac{I_p^2}{(I_*[\alpha])^2}\omega_0,\\
        K ={}& - \frac{3\alpha}{8}\frac{\hbar\omega_0}{L_T(I_*[\alpha])^2}\omega_0,\\
        \xi ={}& -\frac{\alpha}{4}\frac{I_\dc I_p}{(I_*[\alpha])^2}\omega_0e^{-i\varphi_p}.
    \end{align}
\end{subequations}
where once again $\alpha = L_k/L$, $L = L_k + L_g$ and $L_T = lL$ (with $l$ the length of the $\lambda/4$ resonator). 

In Eq.~\ref{eq:Hparam2} we include an implicit dependence of $I_*$ on $\alpha$, since it is not immediately obvious what impact modifying the kinetic inductance fraction has on $I_*$. We know that $I_* \propto I_c$ \cite{Zmuidzinas2012}, thus equivalently we would like to understand how the critical current of the film $I_c$ varies with $\alpha$. 

The kinetic inductance of a thin rectangular wire varies inversely proportional to its width $w$ and thickness $t$ \cite{Annunziata2010a,clem2013inductances}:
\begin{equation}\label{eq:L0}
    L_0 \propto \frac{1}{wt}.
\end{equation}
The thickness dependence can be stronger for very thin films where $t \ll \lambda$ (with $\lambda$ the penetration depth), but is a suitable approximation for the thickness of NbTiN films considered here (i.e. $\geq 50$~nm). To keep our analysis simple, we fix $w$ and only vary the film thickness $t$ to alter $L_k$. For our thin film ($t \ll w$) the geometric inductance of the coplanar waveguide resonator does not depend (or depends only very weakly) on $t$ \cite{clem2013inductances}. Therefore, by keeping $w$ the same we ensure that the geometric properties (inductance and capacitance) remain constant. The kinetic inductance fraction $\alpha$ thus depends on $t$ as:
\begin{equation}
    \alpha = \frac{L_k}{L_k+L_g} = \frac{1}{1+At}.
\end{equation}
where $A = L_g/(L_kt)$ and $L_kt$ is a quantity that does not depend on the film thickness (see Eq.~\ref{eq:L0}). This form of dependence has been observed in experiment over a wide range of NbTiN film thicknesses \cite{kroll2019magnetic}.

In general, thin superconducting films display a complicated relationship between critical current and geometry \cite{hortensius2012critical}. For thin Nb and NbTi films, it has been observed that the critical current density increases for decreasing film thickness \cite{stejic1994effect, pinto2018dimensional}, but decays strongly for extremely thin films where $t \ll \lambda$ \cite{pinto2018dimensional}. Since our film is 50~nm thick and we will be interested in reducing $\alpha$ (i.e. increasing $t$), we assume that $J_c \propto 1/t$. The critical current is straightforwardly related to $J_c$ via $I_c = J_c(wt)$, therefore we assume $I_c$ and consequently $I_*$ to be independent of the film thickness. The pump strength can now be written explicitly as a function of $\alpha$ as:
\begin{equation}\label{eq:xi}
    |\xi| = \frac{\alpha}{4}\frac{I_\dc I_p}{I_*^2}\omega_0.
\end{equation}

The DPA gain is determined by the relative strength of the pump $|\xi|$ and the total resonator bandwidth $\kappa_L = \kappa + \gamma$ (with $\kappa = \omega_0 Q_c^{-1}$ and $\gamma = \omega_0 Q_i^{-1}$), as can be seen from the reflection parameter derived using input-output theory \cite{boutin2017effect,parker2021near}:
\begin{equation}
    \Gamma(\omega) = 
        \frac{\kappa\kappa_L/2 + i\kappa(\Delta + \omega - \omega_p/2)} {\Delta^2 + \big[\kappa_L/2 + i(\omega - \omega_p/2)\big]^2 - |\xi|^2} - 1.
    \label{eqn:S11}
\end{equation}
which is written in the laboratory frame. We observe that at half of the pump frequency ($\omega = \omega_p/2$) and for zero detuning ($\Delta = 0$), the gain increases as $|\xi|^2 \rightarrow \kappa_L^2/4$. 

Assuming that the kinetic inductance fraction $\alpha$ has no impact on the resonator bandwidth $\kappa_L$ (or that any effect can be compensated for by altering the stepped-impedance filter) and that the resonator frequency $\omega_0$ is kept the same (i.e. by changing the length of the resonator), we can maintain a constant gain for different $\alpha$ by adjusting the pump current, making the following substitution in Eq.~\ref{eq:xi}:
\begin{equation}\label{eq:IpIdc}
    I_p \to{} \frac{I_p}{\alpha}.
 \end{equation}
I.e. a smaller kinetic inductance fraction demands a larger pump current, which means a higher pump power. We assume here that the DC bias current remains the same, set at a value below the critical current of the film.

The impedance of the resonator $Z_r$ also depends on $\alpha$:
\begin{equation}\label{eq:Zr}
    \begin{aligned}
        Z_\text{r} ={}& \sqrt{\frac{L}{C}},\\
        ={}& \sqrt{\frac{L_g}{(1-\alpha)C}},\\
        ={}& \frac{Z_{r0}}{\sqrt{(1-\alpha)}}.
    \end{aligned}
\end{equation}
where $Z_{r0} = \sqrt{L_g/C}$ is the resonator impedance in the absence of kinetic inductance.

The pump power is related to the RMS pump current $I_\text{rms} = I_p/\sqrt{2}$ and the resonator impedance according to $P_p = I_\text{rms}^2Z_r$. Inserting Eqs.~\ref{eq:IpIdc} and \ref{eq:Zr}, we find the following dependence of $P_p$ on $\alpha$:
\begin{equation}\label{eq:Pp}
    P_\text{p} = \frac{I_p^2Z_{r0}}{2\alpha^2\sqrt{1-\alpha}}.
\end{equation}
where $I_p$ is the pump current amplitude required in the resonator to achieve a specific level of gain.

Eq.~\ref{eq:Pp} has an optimal point where the pump power is minimized, specifically at $\alpha = 3/4$. This should not be taken too strictly, since for $\alpha \to 1$ ($t \to 0$) the above assumptions (such as the independence of $I_*$ on $t$) are no longer valid. Reducing the kinetic inductance fraction from 0.8 in the current device by a factor of two to 0.4 requires only a modest increase in the pump power of $\Delta P_p = 10\log{[0.8^2\sqrt{1-0.8}/(0.4^2\sqrt{1-0.4})]} = 3.6$~dB. 

\subsection{Signal-to-noise ratio}
Next, we consider how the signal-to-noise ratio (SNR) depends on the kinetic inductance fraction $\alpha$. The SNR is defined by two quantities, the spin sensitivity $N_\text{min}$ of the spectrometer and the total number of spins contributing to the echo $N_\text{tot}$ \cite{morton2018storing}:
\begin{equation}\label{eq:SNR}
    \snr = \frac{N_\text{tot}}{N_\text{min}}.
\end{equation}

Of all the parameters in the theoretical expression for $N_\text{min}$ (Eq.~\ref{eq:Nmin}), only $g_0$ is expected to change with $\alpha$. The coupling strength $g_0 = \delta B_{1\perp}M\gamma_e$ is determined by the electron spin gyromagnetic ratio $\gamma_e$, the spin transition matrix element $M$ and the RMS fluctuations of the magnetic field $\delta B_{1\perp}$. The RMS magnetic field fluctuations are directly proportional to the current in the resonator $\delta I = \omega_0\sqrt{\hbar/(2Z_r)}$ and hence:
\begin{equation}\label{eq:g0_v2}
    \begin{aligned}
        g_0 \propto{}& \frac{\omega_0}{\sqrt{Z_r}}, \\
        ={}& \frac{\omega_0(1-\alpha)^{1/4}}{\sqrt{Z_{r0}}}.
    \end{aligned}
\end{equation}
where we have used Eq.~\ref{eq:Zr} in the second line. This once again provides an $\alpha$ dependence that assumes we alter the kinetic inductance via the film thickness $t$, maintaining the same resonator wire width $w$. We note that Eq.~\ref{eq:g0_v2} makes the underlying assumption that $\delta B_{1\perp}$ only changes with $t$ through its effect on the impedance, which is generally true so long as $t < w\text{,}~\lambda\text{,}~d$ where $d$ is the donor implantation depth (which sets the mean spin-resonator separation), so that the current distribution along the thickness of the film is constant. Finally, we also assume that the resonator frequency $\omega_0$ is kept the same as $\alpha$ is varied by adjusting its length $l$ accordingly.

We arrive at the following $\alpha$ dependence for the spin sensitivity:
\begin{equation}\label{eq:Nminalpha}
    N_\text{min} \propto \frac{1}{(1-\alpha)^{1/4}}.
\end{equation}

We saw in Sec.~\ref{sec:Nd} that the total number of spins is proportional to the effective donor volume $N_\text{tot} \propto V_d$, which in turn is proportional to the length of the last $\lambda/4$ sections in the device. Therefore, we expect $N_\text{tot} \propto l$.

The frequency of the resonator is given by \cite{parker2021near}:
\begin{equation}\label{eq:w0}
    \begin{aligned}
        \omega_0 ={}& \frac{\pi}{2l\sqrt{LC}},\\
        ={}& \frac{\pi\sqrt{1-\alpha}}{2l\sqrt{L_gC}},\\
        ={}& \frac{\pi Z_{r0}\sqrt{1-\alpha}}{2lL_g}.
    \end{aligned}
\end{equation}
rearranging we find:
\begin{equation}\label{eq:w0_v2}
    l = \frac{\pi Z_{r0}\sqrt{1-\alpha}}{2\omega_0L_g}.
\end{equation}
The larger the kinetic inductance fraction, the shorter we must make the resonator to maintain a constant $\omega_0$. This implies that:
\begin{equation}\label{eq:Ntot}
    N_\text{tot} \propto \sqrt{1-\alpha}.
\end{equation}
Finally, putting Eqs.~\ref{eq:Nminalpha} and \ref{eq:Ntot} into Eq.~\ref{eq:SNR} we find:
\begin{equation}\label{eq:SNR_v2}
    \snr \propto (1-\alpha)^{3/4}.
\end{equation}
We reiterate the underlying assumptions in deriving this relation:
\begin{itemize}
    \itemsep-0.5em 
    \item the kinetic inductance fraction $\alpha$ is controlled with the film thickness $t$
    \item the geometric inductance $L_g$ remains constant as $\alpha$ is varied
    \item the frequency $\omega_0$ is kept constant as $\alpha$ is changed by modifying the resonator length $l$
    \item the thickness $t$ changes the spin-resonator coupling strength $g_0$ only through the impedance $Z_r$
\end{itemize}

As expected, increasing the kinetic inductance fraction $\alpha$ has a detrimental effect on the SNR, so this should be minimized, whilst balancing the requirement for low/practical pump powers (Eq.~\ref{eq:Pp}). For a reduction in $\alpha$ from the current value of 0.8 to 0.4, we would expect a factor 2.3 improvement in the SNR. 


Whilst these relations have been found under a number of approximations and assumptions (e.g. $t \ll w$, $t < \lambda$, constant $w$ etc.), improvement in the SNR is likely even for parameter variations outside of these restrictions, though the dependence of $P_p$ and SNR on $\alpha$ should be analyzed further.

\subsection{Sensitivity in the Purcell regime}
A useful measure is the absolute sensitivity, which accounts for the finite measurement repetition time, i.e. $N_\text{min}\times\sqrt{T_1}$. In our device we found $T_1$ to be limited by spin diffusion (Sec.~\ref{sec:T1}), an effect that depends primarily on the pulse sequence being measured and the spin concentration. However, in superconducting microresonators it is possible to reach the so-called ``Purcell regime'', where the spin energy relaxation is mediated by emission of microwave photons into the high-Q resonator \cite{Bienfait2016d}. Here the spin relaxation time depends on the coupling strength and resonator bandwidth:
\begin{equation}\label{eq:T1pur}
    T_1 = \frac{\kappa_L}{4g_0^2}.
\end{equation}
for spins in resonance with the resonator. The absolute sensitivity therefore scales as:
\begin{equation}\label{eq:Nminabs}
    N_\text{min}\sqrt{T_1} \propto \frac{1}{\sqrt{1-\alpha}}.
\end{equation}
resulting in an absolute SNR (i.e. an SNR that takes into account the repetition time) with the dependence:
\begin{equation}\label{eq:SNR_v3}
    \frac{\snr}{\sqrt{T_1}} \propto 1-\alpha.
\end{equation}

\bibliography{references}
\bibliographystyle{unsrtnat}